**Bi-directional personalization reinforcement learning-based architecture with**

**active learning using a multi-model data service for**

**the travel nursing industry**

**By**

**Ezana N. Beyenne**

Ezananb+thesis@gmail.com

Thesis

Submitted in Partial Fulfillment

of the Requirements of the Degree of

MASTER OF SCIENCE IN DATA SCIENCE

March, 2023

Alianna J. Maren, Ph.D., First Reader

Candice Bradely, Ph.D., Second Reader

# Abstract

**Bi-directional personalization reinforcement learning-based architecture with active learning using a multi-model data service for the travel nursing industry**

The challenges of using inadequate online recruitment systems can be addressed with machine learning and software engineering techniques. Bi-directional personalization reinforcement learning-based architecture with active learning can get recruiters to recommend qualified applicants and also enable applicants to receive personalized job recommendations. This paper focuses on how machine learning techniques can enhance the recruitment process in the travel nursing industry by helping speed up data acquisition using a multi-model data service and then providing personalized recommendations using bi-directional reinforcement learning with active learning. This need was especially evident when trying to respond to the overwhelming needs of healthcare facilities during the COVID-19 pandemic. The need for traveling nurses and other healthcare professionals was more evident during the lockdown period.

A data service was architected for job feed processing using an orchestration of natural language processing (NLP) models that synthesize job-related data into a database efficiently and accurately. The multi-model data service provided the data necessary to develop a bi-directional personalization system using reinforcement learning with active learning that could recommend travel nurses and healthcare professionals to recruiters and provide job recommendations to applicants using an internally developed smart match score as a basis.

The bi-directional personalization reinforcement learning based architecture with active learning combines two personalization systems - one that runs forward to recommend qualified candidates for jobs and another that runs backward and recommends jobs for applicants. The reinforcement learning based personalization system uses a custom smart match scoring system to develop a score ranging from 0 (not recommended) to 1 (a perfect match for the job) and then display the top twenty matches to the end user that could be recruiters looking for candidates for each job, or jobs for applicants. The end user then actively gives feedback which the system uses to improve personalized recommendations. The system also uses a batch training process to improve the overall recommendations while also considering other factors, such as fairness.

# Table of Contents

## List of Tables, Illustrations, Figures, or Graphs

# Executive summary

Machine learning and software engineering techniques can help modernize and assist staffing agencies' software systems. For example, machine learning would have significantly improved the recruitment processes in the travel nursing industry during the peak of the COVID-19 pandemic when there were constant changes in staffing and hospital facility needs. With travel nurses becoming more of a fixture in the 21[st] century, hiring nurses for short-term assignments has become a cost-effective method for healthcare facilities to deal with staff shortages. Traveling healthcare professionals (also known as allied professionals) fill short-term positions at healthcare facilities around the nation with temporary openings due to seasonal population fluctuations, maternity leaves, vacations, global pandemics, or other reasons (McGhee 2022).

Healthcare staffing companies help fill shortages in healthcare facilities by using technology for fulfilling jobs as quickly and accurately as possible. The use of machine learning to set up a smart matching system can address this issue. In figure 1 below, each job has nineteen or more matching travelers, and figure 2 shows two hundred ninety-nine to around five hundred seventy-three job matches.

| JobId | TravelerMatchRecommendations |
|---|---|
| C70351A4- | 23 |
| B01B26DA- | 20 |
| F527314A- | 20 |
| 37665630- | 20 |
| 9C343C92- | 19 |
| 29FD6F89 | 19 |

**Figure 1** - Number of traveler matches per available jobs

| ContactID | JobRecommendationPerTraveler |
|---|---|
| 41EE9CC0 | 573 |
| DA6676B3 | 573 |
| 78347760 | 437 |
| 4F025FFE | 413 |
| 5AE9D601 | 309 |

**Figure 2** – Number of open job matches per active traveler

The multitude of choices generated by the smart matching system to choose travelers for jobs or jobs for travelers can be daunting and could lead to decision paralysis. In addition, the large array of options can be discouraging because of the sheer effort spent trying to make the best choices. This is where machine learning-assisted recommendations, especially personalization algorithms, become crucial in helping users narrow their list of choices to be more manageable. In this thesis, two reinforcement learning-based personalization systems are developed to augment the smart matching scoring system by providing users with a more personalized selection of choices in order to improve the submitted numbers of applicants for jobs. Submission numbers are currently a lead indicator for hiring numbers and are tightly correlated.

The two reinforcement learning-based personalization systems formed a bi-directional personalization-based architecture with active learning. The first reinforcement learning model runs forward to recommend qualified candidates for jobs and another that runs backward in an inverse manner and recommends jobs for applicants. The end user then actively gives feedback, which the system uses to improve personalized recommendations. The system also uses a nightly batch training process to improve the overall recommendations while considering other factors, such as fairness.

A reliable bi-directional reinforcement learning-based personalization system requires a data service that processes job feed information in the recommendations. Therefore, a multi-model architecture was developed that contained the orchestration of natural language processing (NLP) models in docker containers and deployed as serverless functions. The multi-model nature of the design enabled each NLP model to be individually designed and deployed as a microservice.

Additionally, a serverless system copied data from the production system to different databases that could be used for training, testing, and validating the models. With this data service infrastructure in place, job processing could be streamlined in near real-time by reducing errors and minimizing manual interventions, which could take anywhere from one hour to two or more weeks. The personalization systems could then take the information from the real-time data service and give personalized recommendations, thereby alleviating the problem of decision paralysis.

Matching qualified applicants to jobs using a personalization system leads to staffing needs being fulfilled as soon as possible and accurately. In addition, this system leads to both parties having their needs met by allowing the healthcare facility to focus on customer-focused healthcare and travelers or allied health professionals to get jobs that fit their needs. As a business within the travel nursing industry evolves, the system will need to adapt to the increase in user engagement. Personalized recommendation systems will eventually improve response time to staffing jobs and increase productivity.

# Chapter 1. Introduction

The staffing and recruitment of travel nurses and allied professionals have become increasingly active in recent years unable to keep up with fast-paced turnover and short-term assignments, particularly during the COVID-19 pandemic. The use of updated technology can help with meeting this increased demand. The challenges of using inadequate online recruitment systems can be addressed with machine learning and software engineering techniques. A bi-directional personalization reinforcement learning-based architecture with active learning runs forward to help recruiters recommend qualified prospective candidates and runs backward to give personalized job recommendations to applicants, hence the bi-directional nature.

Hiring nurses for short-term assignments has become a cost-effective method for healthcare facilities to deal with staff shortages. The travel nursing and allied profession allows healthcare workers to work in different geographic locations that they desire while also gain experience working in different healthcare facilities. In addition, technology has allowed them to work from anywhere thanks to mobile banking, the internet, and mobile phones and the advent of companies specializing in meeting the needs of travelers and healthcare facilities alike.  The demand for traveling nurses and healthcare professionals has been steadily increasing.  U.S Bureau of Labor Statistics projects that by 2025 the nursing staffing shortages could increase to include 500,000 positions (Fastsaff 2014).

 Travel nurses and allied professionals work on short-term assignments typically ranging from 4 to 13 weeks. Their assignments could be extended for up to a year or longer if the traveler decides to accept a full-time position (Fastaff 2014). The fast turnover of staff can create bottlenecks in the recruitment process without the use of technological systems. The use of machine learning can address this problem by helping staffing companies in the travel nursing industry recommend nurses and healthcare professionals while also matching qualified applicants to open jobs in order to fill positions quickly and efficiently. It provides a financial benefit to travelers and healthcare facilities while fulfilling the critical need of helping patients.

A smart match-scoring system was developed to fulfill the ever-changing nature of jobs within the travel nursing industry.. As shown infigures 1 and 2, the multitude of options for matching travelers to jobs and jobs to travelers exposed the smart matching system's deficiencies. These deficiencies caused a condition known as "decision paralysis' because users had a large array of options that tended to discourage them because of the amount of effort they would have to apply to making a decision.

These deficiencies of the smart match scoring system are where recommendations, especially personalization systems, become crucial in helping the recruiter narrow the list of travelers available for each job and the number of jobs available for the travelers. A reinforcement learning-based personalization system with active learning could augment the smart matching scoring system by providing a more personalized selection of choices while allowing users to provide feedback in an active learning fashion.

Figure 3 shows the architecture of reinforcement learning with an active learning-based personalization system, along with a multi-model data service, to feed the personalization system with the necessary data in a real-time manner. A multi-machine learning model data service for processing job information in a real-time also address deficiencies in existing systems. For example, delays could range from a few hours to weeks due to issues ranging from job information not being categorized correctly, requiring human intervention.

The bi-directional personalization reinforcement learning based architecture with active learning combines two personalization systems - one that runs forward to recommend qualified candidates for jobs and another that runs backward and recommends jobs for applicants. The end users get the recommendations and then actively give feedback, which the system uses to improve personalized recommendations.

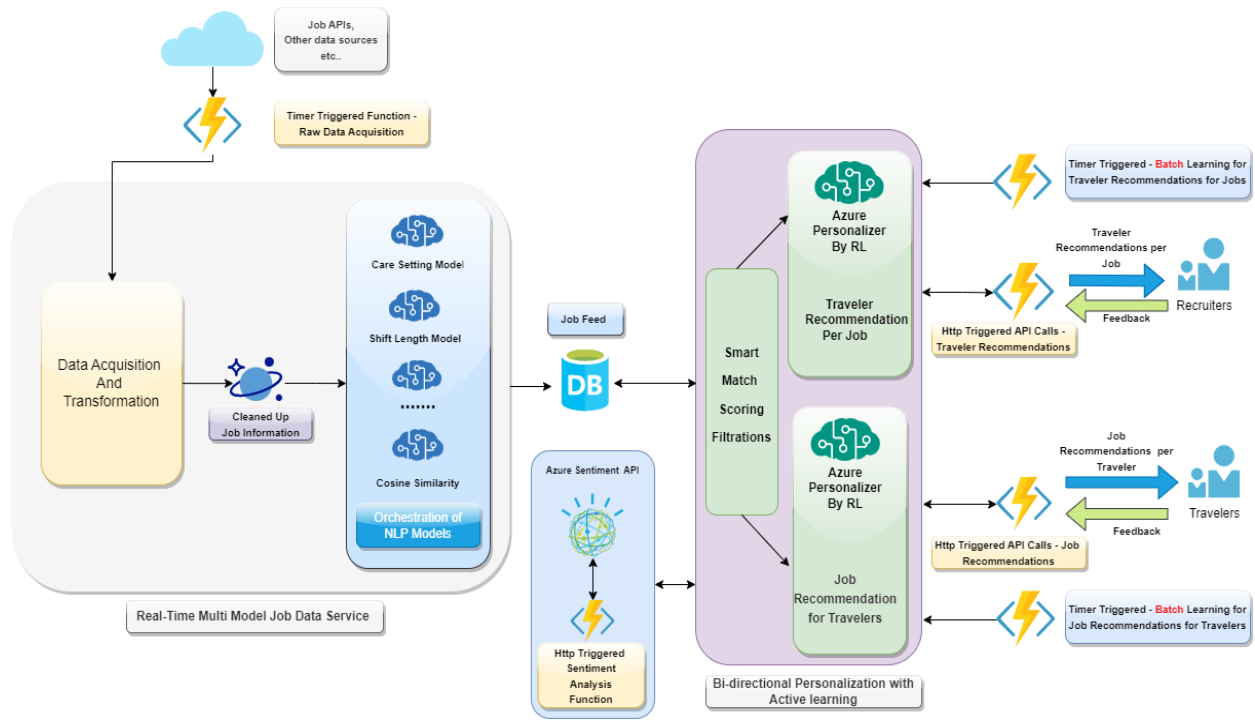

**Figure 3 –** Bi-directional personalization system with multi-model data service architecture

Figure 3 also shows a sentiment analysis and opinion mining service that the smart matching system uses to score travelers' sentiments towards a healthcare facility. The opinion mining portion allows the smart match scoring system to further break down the reviews of the healthcare facility at a more granular level by taking travelers' opinions at the department and healthcare unit levels.

In addition, a batch training methodology could be used to take fairness into account. It would provide a high level view of the data, identifying if travel healthcare professionals are getting most of the jobs or if suitable candidates are left out due to recruiters spending too much time on manual processes or outdated systems.

# Chapter 2. Background

Similar systems should be used as a starting point when deciding to implement a personalization or recommendation system. A taxonomy can be used to analyze recommenders (Falk 2019, 15) and the following dimensions can be used as a guide for the analysis:

- **Domain:** This is used to determine what content needs to be recommended, i.e. a news or movie recommendation system.

- **Purpose:** What is the purpose of the recommendation for both the end user and the company? If it is a staffing company, is the number of correct recommendations that leads to travelers being staffed a measure of how the company is doing? (Falk 2019, 16)

- **Context:** It is the environment in which the user gets the recommendations (Falk 2019, 16-17).

- **Personalization level:** Personalization can come at various levels ranging from ***Non-Personalized:*** which shows the most popular content; ***Semi-Personalized:*** which is based on having some segments belonging to specific memberships; ***Personalized:*** which takes the previous personalization a step further and takes the user's tastes into account. (Falk 2019, 17-18).

## Section 2.1 Traditional Recommendation Systems

Recommender systems have become a crucial component for platforms where users have a myriad of options available. Their business depends heavily on their ability to narrow down the list of options so that users can easily and quickly choose, like finding the right traveler for the job, etc. The traditional types of recommendation systems are as described in the following subsections.

### Section 2.1.1 Collaborative filtering recommendations

Collaborative filtering enables users to be segmented into a particular type because it recommends a list of items for a user based on people with similar tastes (Falk 2019, 183). User preferences are expressed in two categories. **Explicit Rating** is a rating given by a user on a sliding scale, from 0 (disliked) to 5 (extremely liked) (Luo 2021). The system receives explicit ratings as direct feedback from users to show whether they like an item. **Implicit Rating** suggests user choices indirectly by recording views, clicks, purchasing history (Luo 2021).

The sentiment analysis and opinion mining service provided a way for travelers to provide an indirect way to rate healthcare facilities. Figure 4 shows a collaborative filtering algorithm known as the nearest neighborhood algorithm to find similar healthcare facilities for travelers. For example, figure 4 shows that two users have worked at healthcare A and B and had similar sentiment scores for those facilities. Therefore, healthcare C will be recommended to user 1 since user 2 (who has a similar sentiment score) worked there.

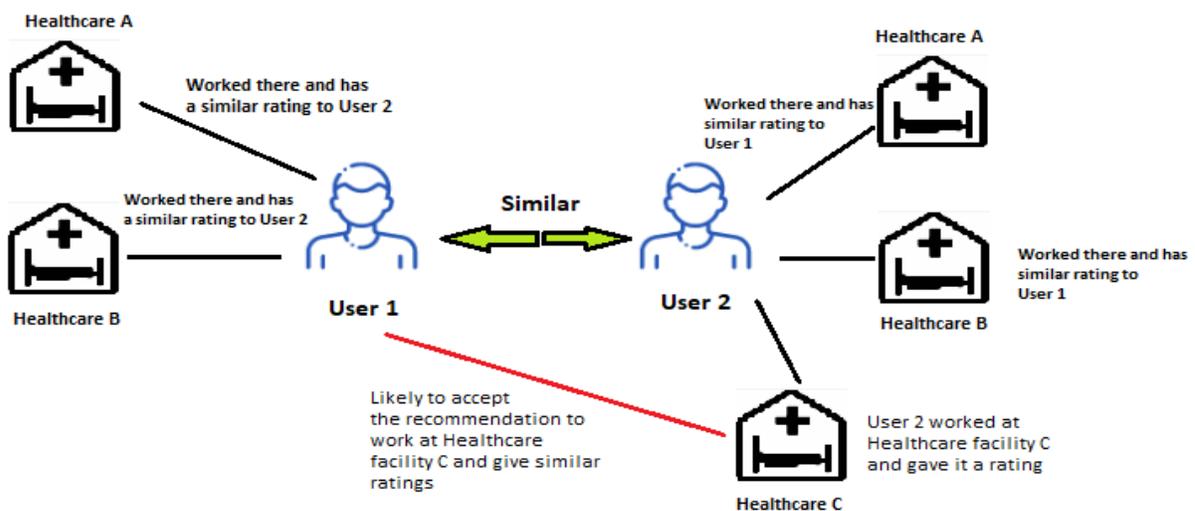

**Figure 4 –** Nearest neighbourhood recommendations using sentiment analysis

We can see in figure 4 a user-based collaborative filtering algorithm was used. An $n$ by $m$ matrix of ratings where traveler $t_i$, where I = 1,..n and healthcare facilities $h_j$, j= 1,..m. The algorithm was then used to predict the rating $r_{ij}$, if the target did not already rate the healthcare facility. The prediction was carried out by calculating the similarities between the target traveler $I$ and all other travelers. Next, the top $N$ similar travelers were selected, and the weighted average ratings from the $N$ travelers with similarities as weights. Figure 5 shows the similarity calculation to predict the rating rij. Usually, when doing similarity calculations, different users give a higher score than others, leading to bias. However, since we were using the sentiment analysis and opinion mining service to give the rating, it was already normalized with a score from zero to one.

$$r_{ij} = \frac{\sum_k \; similarity \; (t_i, t_k) r_{kj}}{number \; of \; ratings}$$

**Figure 5** – Similarity rating prediction calculation

The similarity calculation in the function above is done by a cosine similarity serverless function. It was added to a serverless function because this formula was used to calculate the similarity between the healthcare facility description given and the internal system description described later in the paper. The cosine similarity, as shown in figure 6, calculates the angle between two vectors: the lower the angle between the vectors, the higher the cosine score, yielding a higher similarity factor.

$$cosine \; similarity = \frac{r_i \cdot r_k}{|r_i||r_k|} = \frac{\sum_{j=1}^m r_{ij} r_{kj}}{\sqrt{\sum_{j=1}^m r_{ij}^2 \sum_{j=1}^m r_{kj}^2}}$$

**Figure 6** – Cosine similarity formula

Collaborative filtering has some drawbacks, such as *Sparsity*: the data sets are not dense enough to pick out items that are not popular; *Similarity*: it does not fit recommendations into specific subjects and, as such, is not content-agnostic; *Cold-start problem*: most collaborative-filtering algorithms do not wait for ratings on items to be collected before making recommendations (Ruijt and Bhulai 2021).

### Section 2.1.2 Content based recommendations

Content-based recommendations require much information to extract knowledge from the content. It does not need a user's feedback; instead, it uses information about items' features (Luo 2021). Recommendations can be based on content similarity and get started with two natural language processing (NLP) techniques: Term-Frequency Inverse Document Frequency (TF-IDF) and cosine similarity (Nixon 2022). Appendix A shows a traveler recommender based on job skills that first uses a TF-IDF to obtain a vector representation of the data and then a cosine similarity to define the closest matches.

Content-based recommendations have the advantage of only being concerned about the specific user, making scaling easier. Content-based filtering also captures a user's interests, recommending specific items that a small subset of users are interested in. However, content-based filtering algorithm requires a lot of domain knowledge, and their representations are only as good as the time spent and quality on feature engineering. Moreover, the model cannot expand on the users' current interests.

### Section 2.1.3 Deep learning recommendations

Deep learning recommendations have the added ability to solve complex tasks and deal with complex data because these algorithms can capture non-linear user-item relationships (X. Chen et al. 2021). These algorithms do have some drawbacks when attempting to capture interest dynamics because of the distribution shift (Ruijt and Bhulai 2023). Deep learning techniques use an existing dataset for the training phase which could end up being stale since the real user preferences can undergo a rapid shift (Ruijt and Bhulai 2023).

<p style="text-align:center"><strong>Section 2.2 Deep reinforcement learning techniques</strong></p>

Deep reinforcement learning algorithms address the shortcomings of the deep learning algorithms because they can train an agent to learn from the connected interactions with the environment and combine deep learning and reinforcement learning. The agent of a deep reinforcement learning algorithm can infer dynamic user preferences by actively learning from a user's real-time feedback. Figure 7 shows how deep reinforcement learning algorithms can be divided into model-based and model-free methods. The model-based deep reinforcement learning has an agent that can learn a model of the environment with the aim of estimation function and reward function (X. Chen et al. 2021).

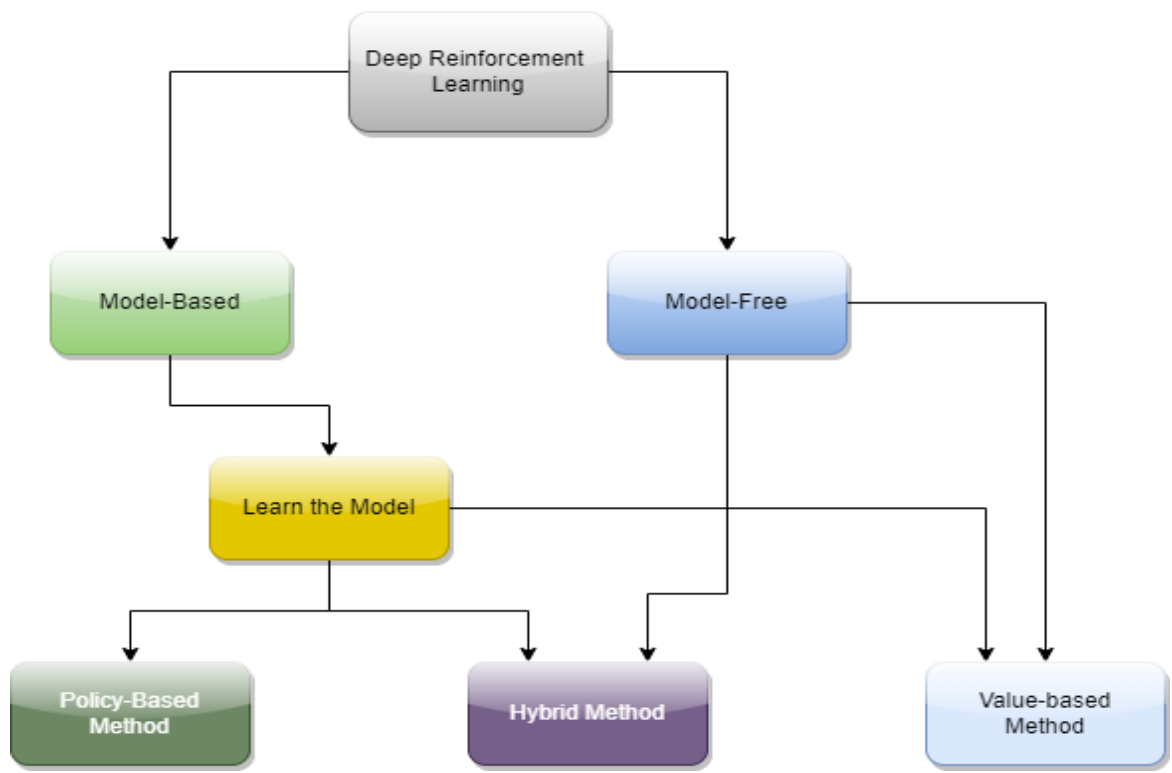

**Figure 7 –** Taxonomy of deep learning for recommender systems (X. Chen et al. 2021)

Value-based, policy-based, and hybrid streams define deep reinforcement learning approaches. When the agent updates the value function to order a policy, they are value-based methods, while the policy-based method learns the policy directly (Ruijt and Bhulai 2023).

Hybrid methods are called actor-critic methods and combine value-based and policy-based methods. Actor-critic has two different networks: the actor, which uses a policy; and the critic, which uses the value-based method intending to evaluate the policy learned by the agent (Ruijt and Bhulai 2023). Deep reinforcement learning balances the learning objective to get the correct data points and could  recommend a sub-optimal recommendation with the aim and belief that it will learn a valuable lesson in the long-run related to its performance (Ruijt and Bhulai 2023).

Various recommendation and personalization algorithms address various needs. For example, deep reinforcement algorithms can allow one-to-one personalization when designed correctly. Therefore, an important consideration for recommendation system design is how these algorithms impact the user experience and how the outputs meet their needs. For example, a contextual bandits reinforcement learning approach used in this thesis was used to determine the best action for a given context across all users to find the maximum average reward (Microsoft Learn 2022e). The contextual bandit learning algorithm is an extension of the multi-armed bandit approach that can test out different actions and automatically learn which action is most rewarding for a given context.

## Chapter 3. Data service architecture for job feed processing

Upon review of the data, a real-time multi-model data service architecture was developed that contained the orchestration of NLP models for the feed processing architecture, along with a testing and deployment infrastructure. Once the data service was in place, the personalization system would operate with certainty using a reliable, testable job feed system. Figure 8 below shows the real-time machine learning-based multi-model job feed data service and the sections of each part of the architecture.

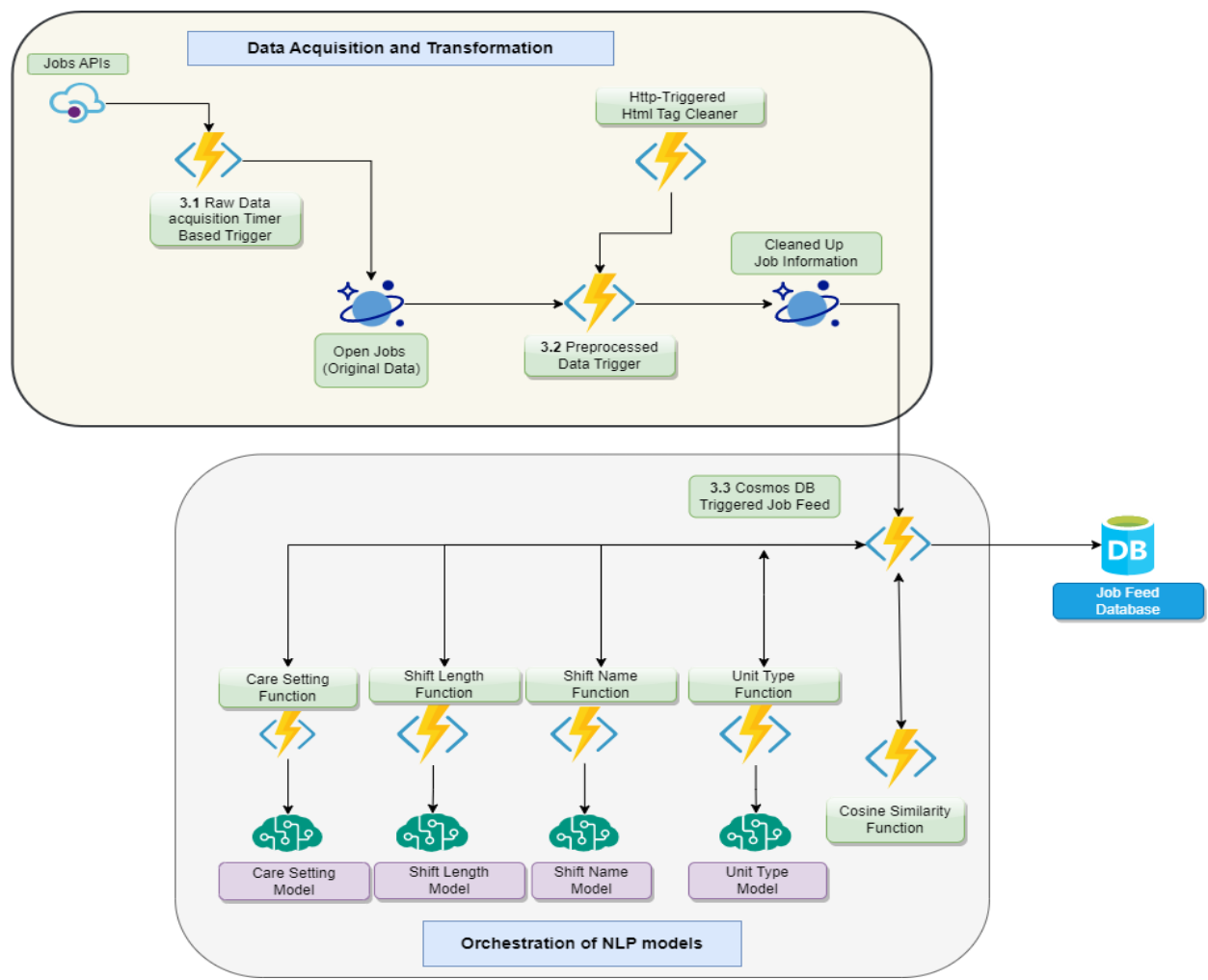

**Figure 8** − Real-time Multi-Model Job Feed Data Service architecture

### Section 3.1 Data Acquistion and Transformation − Raw data trigger

The data acquisition and transformation section in the architecture diagram above shows a serverless function in a docker container that runs on a timer and is responsible for getting data from certain healthcare provider sites' APIs. This function writes the data in the original format in a NoSQL cosmos database for storing open jobs. It automatically scales depending on the workload.

Many authors of job recommendation systems (JRS) state that the vast amount of data they use relies on competition datasets, which are good for training, but rarely for validation because different datasets are designed for different JRS competitions. Their error metrics differ substantially across the different datasets, which raises questions concerning the generalizability of the JRS models trained on one dataset (Ruijt and Bhulai 2023).

The advantage of this system is that disparate job APis are used and normalized. This provides the traveler datasets, jobs datasets, hospital facilities datasets, and interaction data, which helps resolve language inconsistencies between travelers and recruiters that could be more troublesome regarding some content-based and knowledge-based JRS systems (Ruijt and Bhulai 2023).

### Section 3.2 Data Acquistion and Transformation – Preprocessed Data trigger

The preprocessed data function wakes up when open jobs are written to the open jobs cosmos db. This containerized serverless function calls another HTTP-triggered function that cleans up HTML tags etc., using the Beautiful Soup python library. It then writes the cleaned-up job information to another Cosmos database.

### Section 3.3 Orchestration of NLP Models – DB Triggered Job Feed functionality

Once data has been written to the job information NoSQL database, a Cosmos DB-triggered serverless function can pick up each job and call various machine learning-based serverless functions that are responsible for different mapping features (such as shift length, hospital unit type, hospital care settings) to a column in the job feed data transfer object before finally saving it into the job feed database. The machine learning models consist of four named entity recognition serverless functions and a cosine similarity serverless function that are independently trainable, testable, and deployable.

### 3.3.1 Orchestration of NLP Models – Cosine similarity model

Many misspellings from the job APIs were identified during the data analysis phase. The serverless function can match names, regardless of manual misspellings and mistakes, using a correct naming convention. Anything below 70 percent was written to a table, and a person was notified to assist in deciphering the name.

### 3.3.2 Orchestration of NLP Models – Named Entity Recognition models

Four Named Entity Recognition (NER) models were developed to identify critical features using Microsoft's Language Understanding Intelligent Service (LUIS). These four features are hospital care setting, shift length, shift name, and hospital unit. LUIS was used after spending time for proof of concept developing NER models using Spacy, then Keras and LTSM models. Additionally, the NER models and systems need to handle the amount of data coming through and have a UI to quickly add new entities and identify them in the desired sentences.

After going through different providers, LUIS from Microsoft was chosen since it provided all requirements and was available in Azure. Figure 9 below shows the UI for developing the CareSetting named entity recognition (NER). In Figure 10, the UI was used for developing Shift Name NER.

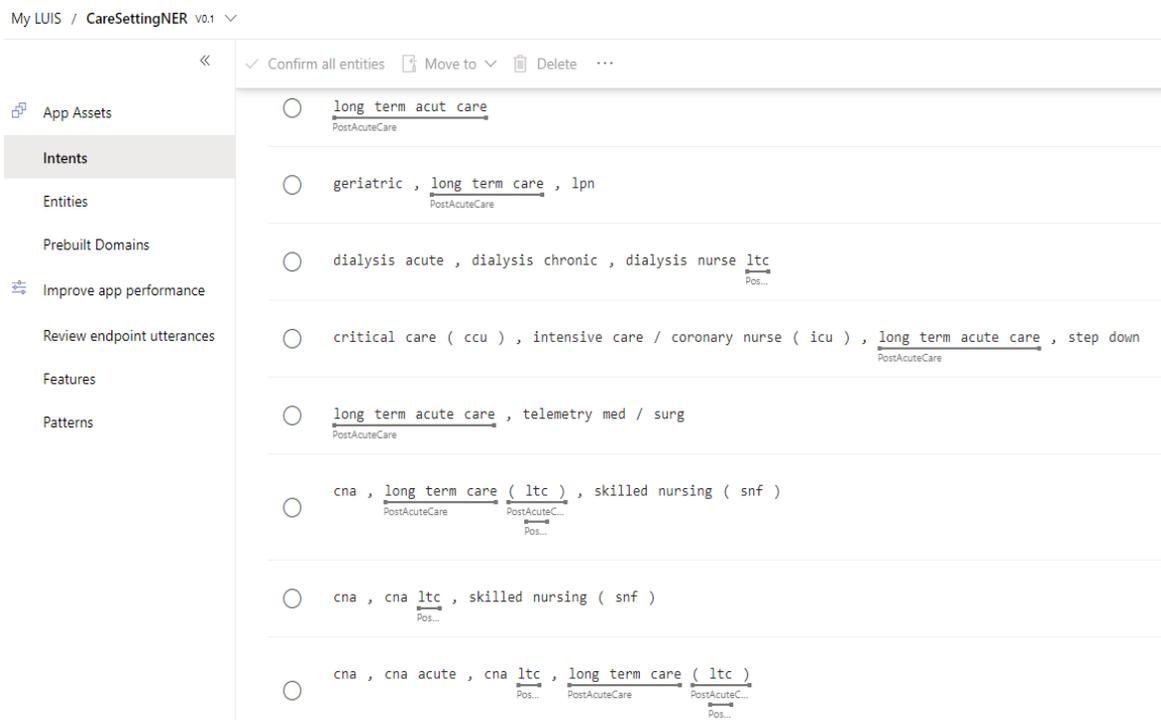

**Figure 9 –** UI used for developing CareSetting Named Entiry Recognition (NER)

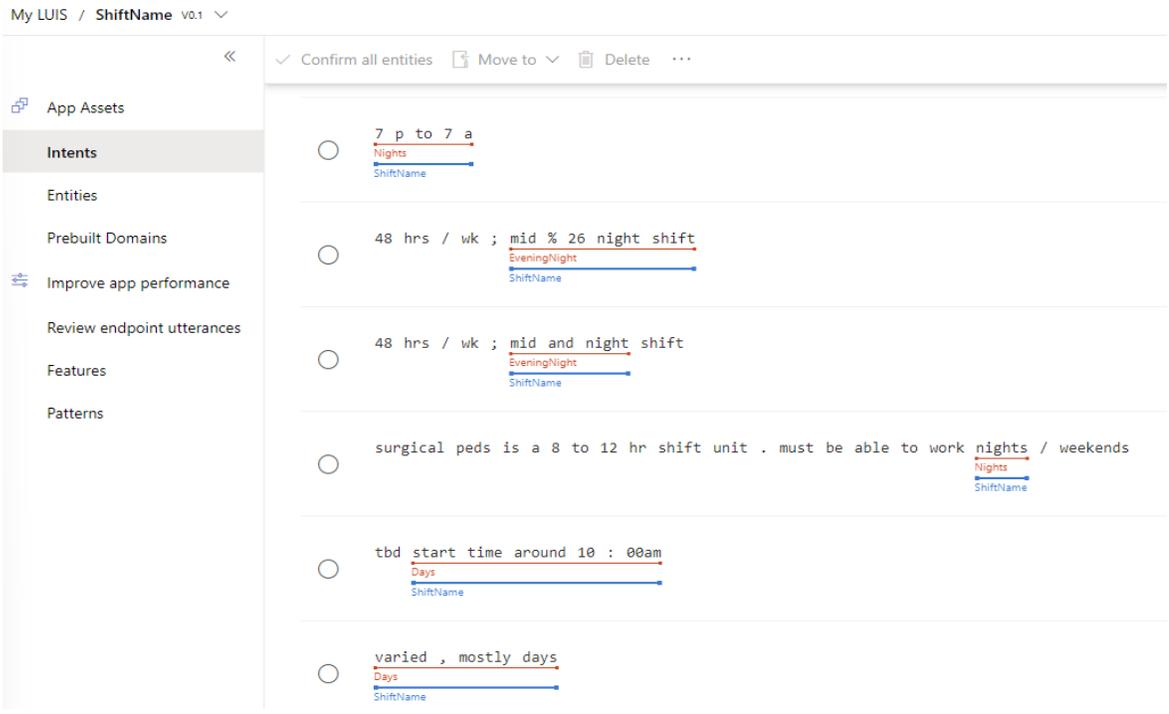

**Figure 10 –** UI used for developing Shift Name NER

One of the strengths of NER models is that they often correctly identified instances that produced multiple correct answers, but a decision was made to choose one. For instance, in Figure 11, the Unit Type NER model correctly identified two correct unit types, but the first one was chosen.

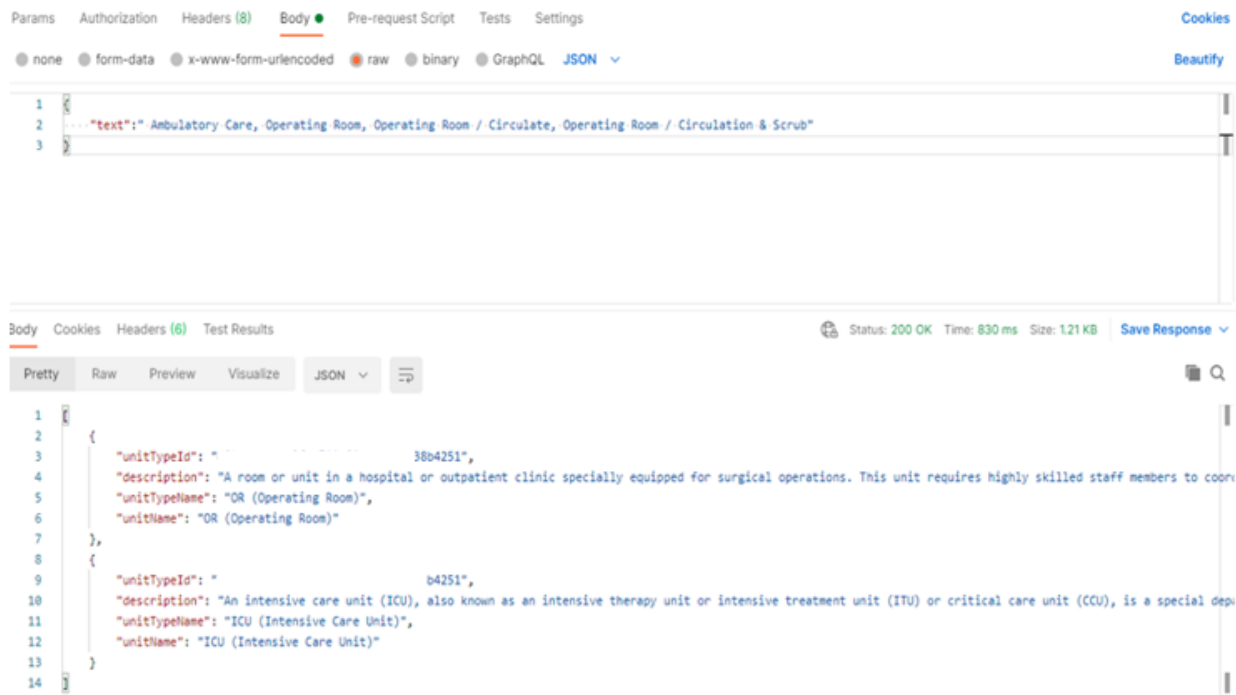

**Figure 11 –** Multiple Unit Types selected (OR and ICU) by NER model

### Section 3.4 Testing and Model validation framework

The black-box nature of machine learning systems incites little confidence. Therefore, a testing framework was developed to ensure that the machine learning systems perform as expected, including unit and integration tests that must pass before the serverless functions are deployed in each environment.

### 3.4.1 Replicating Job Data from Production

Figure 12 shows containerized serverless functions that are used to copy data from production, then scrub the data and copy the information to other databases continuously. This production data was then used to validate the output of the models by running integration tests, addressing data and model drifts.

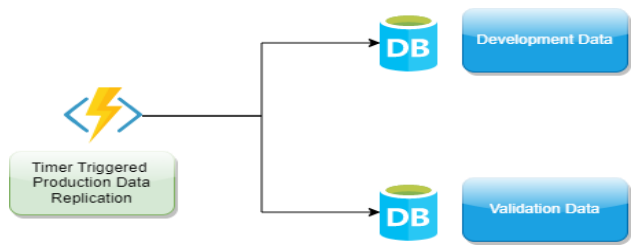

**Figure 12** – Timer triggered data replication process

### 3.4.2 Multi-Model Data Service vs. existing architecture

Figure 13 below shows the differences between the multi-model data service and the existing system. In the first column Diff_Of_Date_Detection, the figure shows the difference between when the multi-model first detected the job and when the current system detected the job. The differences in this sample ranged from six hours to around nine hours. The second highlighted column Diff_of_Creation_Date shows the time difference in hours between when the multi-model and the current system processed the job. For example, the figure below shows that the multi-model data service creates and processes the various jobs about five hours and twenty-nine minutes to eight hours and eight minutes faster than the existing system.

| ExternalIdentifer | Date_Model_First_Detected_Job | Date_Current_Job_System_Detected_Job | Diff_Of_Date_Detection | Model_Created_Job_Date | Current_System_Created_Date | Diff_of_Creation_Date |
|---|---|---|---|---|---|---|
| 199298 | 2021-10-18 10:30:00.2333333 | 2021-10-18 19:30:00.3800000 | 09:00:00 | 2021-10-18 10:00:11.6898187 | 2021-10-18 18:08:11.6566667 | 08:08:00 |
| 199298 | 2021-10-18 11:00:00.3600000 | 2021-10-18 19:30:00.3800000 | 08:30:00 | 2021-10-18 10:00:11.6898187 | 2021-10-18 18:08:11.6566667 | 08:08:00 |
| 220566 | 2022-05-13 14:30:00.2800000 | 2022-05-13 22:30:00.4633333 | 08:00:00 | 2022-05-13 14:00:12.7445120 | 2022-05-13 21:02:21.8800000 | 07:02:09 |
| 199298 | 2021-10-18 11:30:00.3800000 | 2021-10-18 19:30:00.3800000 | 08:00:00 | 2021-10-18 10:00:11.6898187 | 2021-10-18 18:08:11.6566667 | 08:08:00 |
| 220566 | 2022-05-13 15:00:00.6533333 | 2022-05-13 22:30:00.4633333 | 07:30:00 | 2022-05-13 14:00:12.7445120 | 2022-05-13 21:02:21.8800000 | 07:02:09 |
| 199299 | 2021-10-18 12:00:00.2800000 | 2021-10-18 19:30:00.3800000 | 07:30:00 | 2021-10-18 11:30:05.6207766 | 2021-10-18 18:09:14.8233333 | 06:39:09 |
| 199298 | 2021-10-18 12:00:00.2800000 | 2021-10-18 19:30:00.3800000 | 07:30:00 | 2021-10-18 10:00:11.6898187 | 2021-10-18 18:08:11.6566667 | 08:08:00 |
| 220566 | 2022-05-13 15:30:00.3066667 | 2022-05-13 22:30:00.4633333 | 07:00:00 | 2022-05-13 14:00:12.7445120 | 2022-05-13 21:02:21.8800000 | 07:02:09 |
| 199314 | 2021-10-18 15:30:00.2800000 | 2021-10-18 22:30:00.4400000 | 07:00:00 | 2021-10-18 15:00:18.0965766 | 2021-10-18 21:39:05.8133333 | 06:38:47 |
| 199299 | 2021-10-18 12:30:00.3000000 | 2021-10-18 19:30:00.3800000 | 07:00:00 | 2021-10-18 11:30:05.6207766 | 2021-10-18 18:09:14.8233333 | 06:39:09 |
| 199298 | 2021-10-18 12:30:00.3000000 | 2021-10-18 19:30:00.3800000 | 07:00:00 | 2021-10-18 10:00:11.6898187 | 2021-10-18 18:08:11.6566667 | 08:08:00 |
| 233861 | 2022-10-06 15:00:00.7100000 | 2022-10-06 21:30:00.8466667 | 06:30:00 | 2022-10-06 14:31:52.9313315 | 2022-10-06 20:32:21.9233333 | 06:00:29 |
| 220566 | 2022-05-13 16:00:00.4366667 | 2022-05-13 22:30:00.4633333 | 06:30:00 | 2022-05-13 14:00:12.7445120 | 2022-05-13 21:02:21.8800000 | 07:02:09 |
| 199314 | 2021-10-18 16:00:00.4733333 | 2021-10-18 22:30:00.4400000 | 06:30:00 | 2021-10-18 15:00:18.0965766 | 2021-10-18 21:39:05.8133333 | 06:38:47 |
| 199299 | 2021-10-18 13:00:00.4533333 | 2021-10-18 19:30:00.3800000 | 06:30:00 | 2021-10-18 11:30:05.6207766 | 2021-10-18 18:09:14.8233333 | 06:39:09 |
| 199302 | 2021-10-18 13:00:00.4533333 | 2021-10-18 19:30:00.3800000 | 06:30:00 | 2021-10-18 12:30:10.2557676 | 2021-10-18 18:29:43.8833333 | 05:59:33 |
| 199301 | 2021-10-18 13:00:00.4533333 | 2021-10-18 19:30:00.3800000 | 06:30:00 | 2021-10-18 12:30:12.1907386 | 2021-10-18 18:28:54.7433333 | 05:58:42 |
| 199298 | 2021-10-18 13:00:00.4533333 | 2021-10-18 19:30:00.3800000 | 06:30:00 | 2021-10-18 10:00:11.6898187 | 2021-10-18 18:08:11.6566667 | 08:08:00 |
| 233861 | 2022-10-06 15:30:00.2833333 | 2022-10-06 21:30:00.8466667 | 06:00:00 | 2022-10-06 14:31:52.9313315 | 2022-10-06 20:32:21.9233333 | 06:00:29 |
| 233872 | 2022-10-06 15:30:00.2833333 | 2022-10-06 21:30:00.8466667 | 06:00:00 | 2022-10-06 15:03:11.2894100 | 2022-10-06 20:32:50.9466667 | 05:29:39 |

**Figure 13** – Time differences between service and the existing system

Figure 14 shows a sample of the service that has jobs mapped and errors within the existing system that causes delays in retrieving the job data, and these delays can range from a few hours to days or more. For example, healthcare facilities need to have their unit types correctly mapped.

| ExternalIdentifer | StartingRunDate | ModelCreatedDate | JobFeedCreatedDate | ModelModifiedDate | JobFeedModifiedDate |
|---|---|---|---|---|---|
| 186874 | 2021-07-12 23:10:00.4866667 | 2021-07-12 13:00:16.4330302 | NULL | 2021-07-12 23:02:38.1793509 | NULL |
| 186875 | 2021-07-12 23:10:00.4866667 | 2021-07-12 13:00:17.2769400 | NULL | 2021-07-12 23:00:08.3454746 | NULL |
| 186877 | 2021-07-12 23:10:00.4866667 | 2021-07-12 14:00:19.0180574 | NULL | 2021-07-12 23:00:37.0173127 | NULL |
| 186879 | 2021-07-12 23:10:00.4866667 | 2021-07-12 14:00:19.6272460 | NULL | 2021-07-12 23:00:16.8770853 | NULL |
| 186878 | 2021-07-12 23:10:00.4866667 | 2021-07-12 14:00:21.8756978 | NULL | 2021-07-12 23:00:09.0366212 | NULL |
| 186899 | 2021-07-12 23:10:00.4866667 | 2021-07-12 15:00:16.5557778 | NULL | 2021-07-12 23:01:56.2857858 | NULL |
| 186893 | 2021-07-12 23:10:00.4866667 | 2021-07-12 15:00:18.0856826 | NULL | 2021-07-12 23:02:07.7610424 | NULL |
| 186885 | 2021-07-12 23:10:00.4866667 | 2021-07-12 15:00:19.2191632 | NULL | 2021-07-12 23:02:41.1676322 | NULL |
| 186892 | 2021-07-12 23:10:00.4866667 | 2021-07-12 15:00:22.1051200 | NULL | 2021-07-12 23:02:18.5267089 | NULL |
| 186889 | 2021-07-12 23:10:00.4866667 | 2021-07-12 15:00:27.2277043 | NULL | 2021-07-12 23:00:14.0885809 | NULL |
| 186894 | 2021-07-12 23:10:00.4866667 | 2021-07-12 15:00:27.8643223 | NULL | 2021-07-12 23:02:14.2231765 | NULL |
| 186891 | 2021-07-12 23:10:00.4866667 | 2021-07-12 15:00:30.6782705 | NULL | 2021-07-12 23:02:29.2049015 | NULL |

**Figure 14 –** ML service has jobs mapped, but current system does not have jobs showing

## Chapter 4. Bi-directional personalization system architecture

Once the data from the job feed processing system was addressed, the next phase of the architecture is the development of a personalized recommendation system that recommends travelers for active jobs and its inverse recommending jobs for travelers while having the ability to actively learn in real-time and address concerns from stakeholders. A review of traditional recommendation techniques included collaborative-filtering and content-based recommendation as well as deep learning and reinforcement learning techniques.

After looking at the various recommendation or filtering algorithms mentioned above, the smart matching scoring system was developed as a semi-personalized filtering system. A smart match scoring system aims to see job submission rates jump. The reason is that submission numbers are a leading indicator for hiring numbers. Job submission rates are tightly correlated to hiring numbers. Once the smart matching scoring system was in place, the need for a personalized recommendation system was evident because each active job could have anywhere from one to fifty or more recommendations of travelers. Travelers could also receive over a hundred job recommendations. The probability of a large number of recommendations of travelers for each job or job recommendation for travelers could lead to "decision paralysis". By using reinforcement learning, the bi-directional personalization system needed to infer dynamic preferences in real-time in an automated, scalable, and adaptable fashion.

Azure personalizer was chosen since it provided reinforcement learning with an active learning system and was part of the Azure family of Cognitive Services lineup, which meant easier integration with Azure, including Azure Serverless Functions and Azure DevOps. In addition, two reinforcement learning models were required to address the inverse needs of traveler recommendations for active jobs and job recommendations for travelers. The result is a bi-directional personalizer system that combines two personalization systems – one that runs forward to recommend qualified travelers for jobs and another that runs backward and recommends jobs for travelers.

Two serverless functions per personalization models were developed to interact with the two reinforcement learning personalization systems in an active and batch learning fashion. The batch learning timer-triggered endpoints would help train the two bi-directional models using the total smart match score as a reward value and also helped initialize the two personalization models to avoid the "Cold Start" problem when the models are first starting. The active learning HTTP-triggered endpoints would take the users' probability scores as reward values.

In addition, sentiment analysis with opinion mining was used to take traveler feedback in the smart match scoring system that acted as a collaborative filtering health facility recommendation system and was used as a hospital sentiment scoring method.

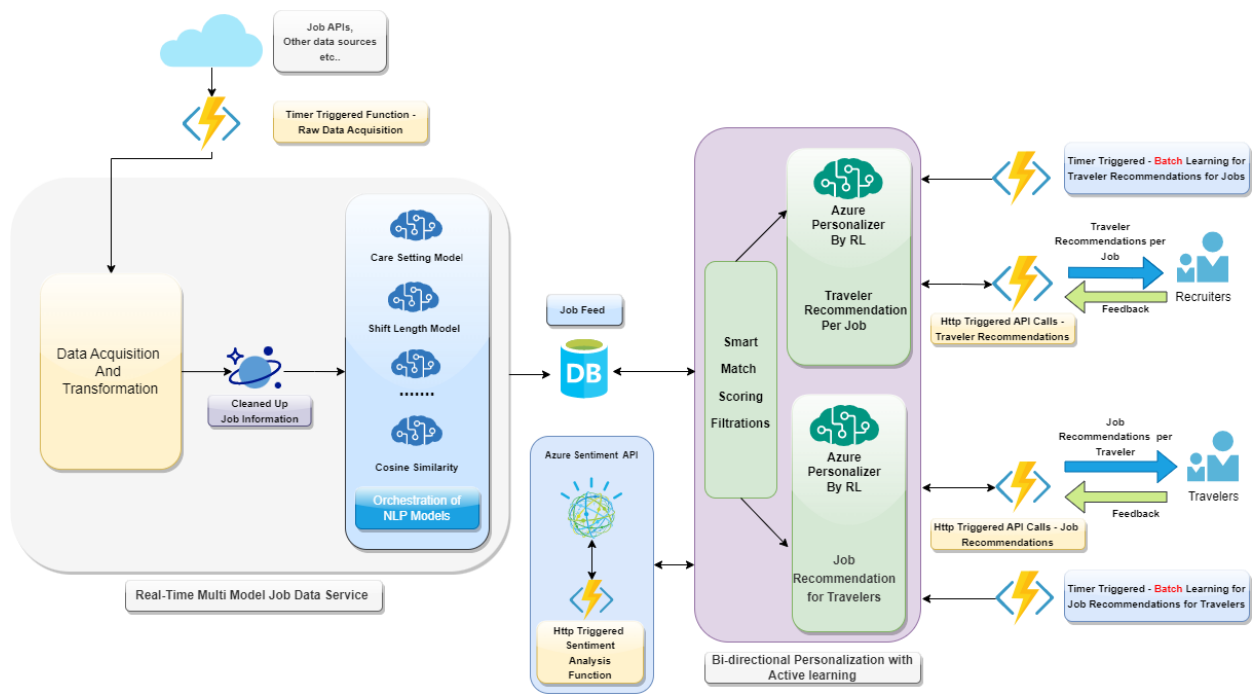

**Figure 15 –** Personalized bi-directional personalizer architecture with multi-model data service

### Section 4.1 Smart match scoring filtering method

Features that comprised active jobs and suitable travelers need to be identitfied to develop the personalization systems. To identify the jobs, the hospital facility  must meet the following characteristics:

1. The hospital should be active.

2. The hospital needed to be in Ok status and not a financial risk or have a caution status.

3. The account status should not be a prospect or duplicate account.

4. The job status should be open.

5. The start date should be within range.

To identify the traveler, the traveler had to meet the following characteristics:

1. Should be in 'ok to use' status.

2. The status code should be active.

3. The traveler should be available.

4. The Traveler status should be in good standing.

Once the job and traveler criteria were identified, a similarity matching and scoring system was developed. The scoring system ranged from 1.0 to 0, so the active learning reward system of the personalizer system could use it. The similarity matching and scoring system is shown below:

1. **Availability Date:** If the "*Availability Date*" is between 0 and 30 days of the job's "*Start Date*", give it 1 point. If the "Availability Date" is between 30 and 60 days from the job's "*Start Date*", then give it 0.8 points, otherwise 0 points.

2. **Desired State:** If the job is in the traveler's "Desired state, "it gets 1 point. Otherwise, if the traveler has not given a "Desired state," it is worth 0.8 points.

3. **Primary or Secondary Skills:** If the "Primary Skill" matches the job's skill, give it 1 point. If the traveler's secondary skill matches the job skill, then 0.8 points, otherwise 0 points.

4. **Desired Shift:** If the traveler's desired Shift information matches the job's shift information, give it 1 point. If the traveler has not given a desired shift, then 0.8 points.

5. **Licensed States:** If "*Licensed States*" match 1 point. If no "*Licensed States*" is entered, 0.8 points, otherwise 0 points.

6. **Certification:** The range is 1 match between the job and the traveler; otherwise, 0.8 points.

7. **Traveler Status:** It can range from 1 for the current status to 0.5 for those interested in travel.

8. **Hospital Sentiment:** This uses the sentiment that travelers who have worked there give. Opinion mining further breaks it down into hospital sentiment, hospital units, etc.

9. **Hospital Bed Size:** Desired bed size by the traveler.

10. **Take-Home Pay:** 1.0 if the take-home pay is greater than or equal to the requested pay. Otherwise, 0.8 was used.

11. **Previous Assignment:** has a similar scoring system for previous assignments.

12. **Client Level:** has a scoring system for the client level.

### Section 4.2 Sentiment Analysis and Opinion mining function

Since hospital sentiment was one of the matching features, Azure's sentiment analysis was added to a containerized serverless function, and the traveler's review of the hospital facilities and hospital unit was considered. Travelers could have a good opinion of the hospital unit but not a great opinion of the hospital facility. The sentiment analysis system had to consider their verbose input and decipher if it was ok to get a recommendation. Figure 16 shows a sample review of a healthcare facility by a traveler, and the sentiment analysis output.

The sentiment analysis and opinion mining helps travelers and the staffing company to see how other applicants and staffing members have positively interacted with the hospital facilities. Sentiment analysis and opinion mining can add collaborative filtering with the idea that candidates with similar past experiences will probably want to work at a facility or not. Additionally, it allows both the staffing company and other candidates to trust the real feedback travelers have made instead of just relying on smart matching techniques, which, most times, is an estimated guess (Chen et al. 2021).

"Text": "Nice work environment! I had a great unobstructed view of the        Hospital campus but Oncology department was old and the toilet in disrepair. I will recommend it for the colleagues, otherwise, might be better to work at a hospital with the department that has not been neglected.",
"IncludeOpinionMining": "False"

"overallSentiment": 3,
"overallSentimentName": "Mixed",
"overallConfidenceScores": {
    "positive": 0.68,
    "neutral": 0.0,
    "negative": 0.32
},
"sentences": [
    {
        "sentiment": 0,
        "sentimentName": "Positive",
        "text": "Nice work environment!",
        "confidenceScores": {
            "positive": 1.0,
            "neutral": 0.0,
            "negative": 0.0
        },
        "opinions": []
    },
    {
        "sentiment": 2,
        "sentimentName": "Negative",
        "text": "I had a great unobstructed view of the        Hospital campus but Oncology department was old and the toilet in disrepair.",

**Figure 16** – Sentiment analysis of a traveler's hospital review from text

**Section 4.3 Data acquisition**

The data was written to a SQL server database, and stored procs were then used to get the data if a user requested it or in a batch format for the personalization model to train and then deploy the system if it passed tests. An example of a matching and scoring system is shown below in Figure 17 and Figure 18. For example, figure 17 shows traveler recommendations for a job, and figure 18 shows a job recommendation for a traveler.

| JobId | JobNumber | StartDate | PostionOpen | JobTypeName | NumberOfBeds | LicenseToSubmitName | Certifications | BillRate | HospitalName | City | ShiftLengthName | ShiftName | StateName | JobStatus | JobStatusName | GuaranteedHoursName |
|---|---|---|---|---|---|---|---|---|---|---|---|---|---|---|---|---|
| D0EC613A | 216891 | 2021-09-15 | 2 | Core | 350 | Yes | BLS | 165.00 | Hospital | | 12 | Nights | FL | 717660000 | Open | 36 |

| TotalSmartMatchScore | JobId | JobNumber | PostionOpen | TravelerStatus | TravelerStatusName | DesiredTravelerStatusScore | ContactId | FullName | StartDate | AvailabilityDate | AvailabilityDateDiff | Availabilit |
|---|---|---|---|---|---|---|---|---|---|---|---|---|
| 0.940000 | D0EC613A | 216891 | 2 | 717660002 | Current MS Employee | 1.0 | 9D230FEF | | 2021-09-15 | 2021-09-06 00:00:00.000 | 9 | 1.0 |
| 0.920000 | D0EC613A. | 216891 | 2 | 717660002 | Current MS Employee | 1.0 | 29692A5- | | 2021-09-15 | 2021-09-04 00:00:00.000 | 11 | 1.0 |
| 0.920000 | D0EC613A- | 216891 | 2 | 717660002 | Current MS Employee | 1.0 | D70ED01D | | 2021-09-15 | 2021-09-04 00:00:00.000 | 11 | 1.0 |
| 0.920000 | D0EC613A | 216891 | 2 | 717660002 | Current MS Employee | 1.0 | 2C3183D2 | | 2021-09-15 | 2021-09-05 00:00:00.000 | 10 | 1.0 |
| 0.920000 | D0EC613A-F | 216891 | 2 | 717660004 | Current Traveler | 1.0 | F2C04B45- | | 2021-09-15 | 2021-08-16 00:00:00.000 | 30 | 1.0 |
| 0.920000 | D0EC613A | 216891 | 2 | 717660005 | Pending MS Employee | 0.8 | 386F536F | | 2021-09-15 | 2021-09-06 00:00:00.000 | 9 | 1.0 |
| 0.910000 | D0EC613A. | 216891 | 2 | 717660014 | Previous MS Employee | 0.7 | E651EBAB- | | 2021-09-15 | 2021-08-23 00:00:00.000 | 23 | 1.0 |
| 0.900000 | D0EC613 | 216891 | 2 | 717660002 | Current MS Employee | 1.0 | 9F9017B6- | | 2021-09-15 | 2021-08-16 00:00:00.000 | 30 | 1.0 |
| 0.900000 | D0EC613A. | 216891 | 2 | 717660002 | Current MS Employee | 1.0 | 7E467A1D | | 2021-09-15 | 2021-09-13 00:00:00.000 | 2 | 1.0 |

**Figure 17** – Traveler matches to a job by smart matching score

| ContactId | | AvailabilityDate | LicensedStates | PreferedHospitalSize | TakeHomePay | DesiredStates | FullName | Certifications | PrimarySkillName | SecondarySkill | ShiftLengthName | ShiftName |
|---|---|---|---|---|---|---|---|---|---|---|---|---|
| | 1 | 2021-09-13 00:00:00.000 | NULL | NULL | 1450.0000 | NULL | Charlene Crawford | NULL | Phlebotomy | NULL | All | All |

| TotalSmartMatchScore | ContactId | FullName | JobId | JobNumber | PositionOpen | TravelerStatus | TravelerStatusName | DesiredTravelerStatusScore | StartDate |
|---|---|---|---|---|---|---|---|---|---|
| 0.880000 | | | | 21605 | 1 | 717660000 | Unknown | 0.5 | 2021-09-13 |
| 0.870000 | | | | 23226 | 1 | 717660000 | Unknown | 0.5 | 2021-09-14 |
| 0.880000 | | | | 23964 | 1 | 717660000 | Unknown | 0.5 | 2021-09-24 |
| 0.870000 | | | | 21557 | 3 | 717660000 | Unknown | 0.5 | 2021-10-04 |
| 0.880000 | | | | 23961 | 1 | 717660000 | Unknown | 0.5 | 2021-10-11 |
| 0.880000 | | | | 21730 | 1 | 717660000 | Unknown | 0.5 | 2021-10-12 |
| 0.860000 | | | | 229933 | 1 | 717660000 | Unknown | 0.5 | 2021-10-18 |
| 0.860000 | | | | 22993 | 1 | 717660000 | Unknown | 0.5 | 2021-10-18 |
| 0.860000 | | | | 23417 | 2 | 717660000 | Unknown | 0.5 | 2021-10-19 |

**Figure 18 –** Job matches to a traveler by smart matching score

### 4.3.1 Data acquisition for traveler recommendations for active jobs

Figure 19 shows the tables, view, and smart match scoring, also known as similarity scoring, that feeds the Azure personalizer's reinforcement learning system for matching travelers to jobs. This data is used for both the active learning API calls and the batch learning system.

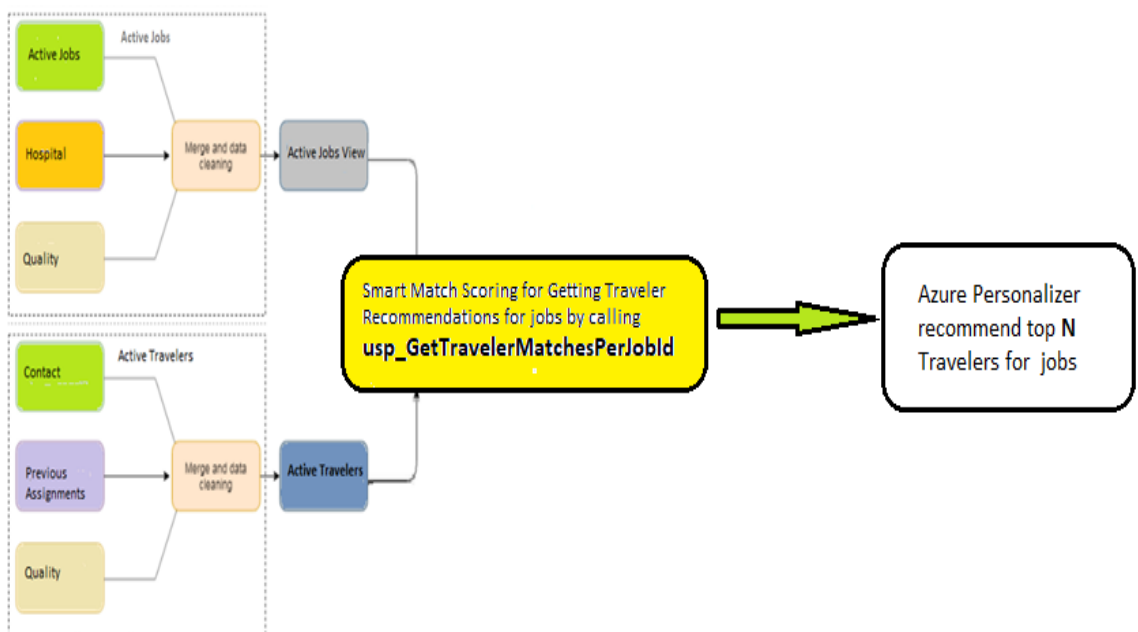

**Figure 19 –** Data acquisition ror traveler recommendations by jobs

### 4.3.2 Data acquisition for job recommendations for traveler preferences

Figure 20 shows the tables, view, and similarity scoring, that feeds the azure personalizer's reinforcement learning system for matching active jobs to travelers and is used for both the active learning API calls and the batch learning system.

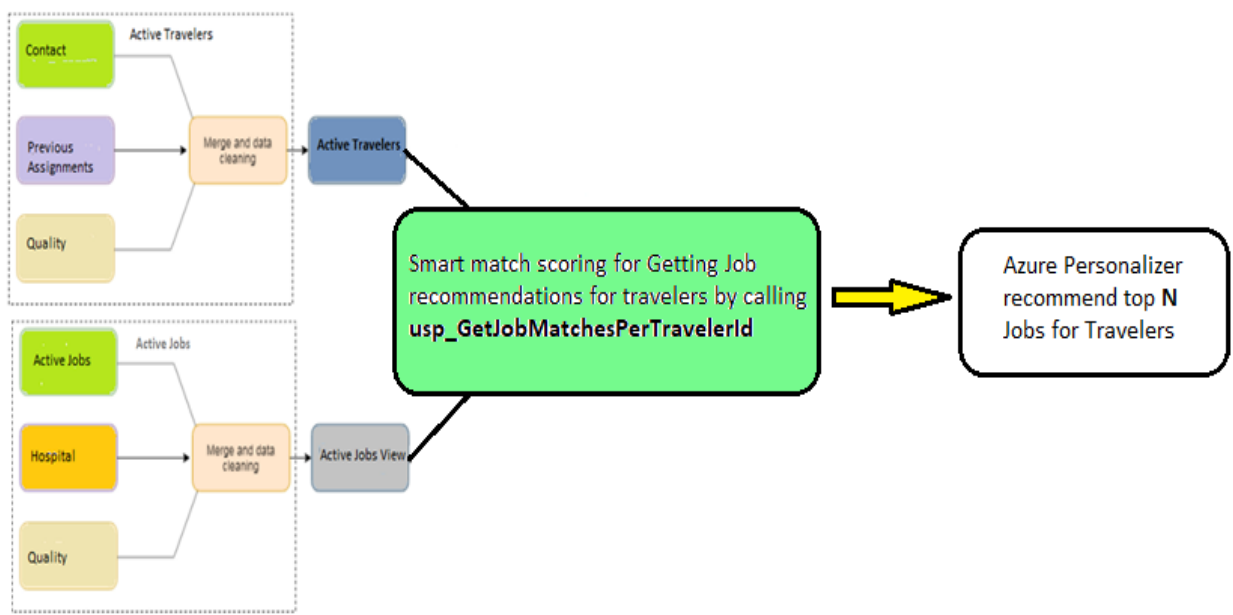

**Figure 20** – Data acquisition for job recommendations by traveler

### Section 4.4 Bi-directional personalization system code

Traveler and job recommendation systems have more challenges than a current or single-step bandit-based approach. The additional challenges are as follows:

- The action space is constantly evolving. Many jobs are flowing in, with unique characteristics such COVID-19 new laws, licenses, and certification needs.

- Traveler preferences are dynamic, too, that can evolve. They might get new training, licenses, or certifications, in addition to their traveling preferences changing over time and their desire to work at other healthcare facilities, states, etc.

- Job or traveler recommendations are a multi-step problem. For instance, multiple jobs can be matched to a traveler, and/or conditions such as a pandemic might lead travelers to be overwhelmed.

- The agent only sees a limited observation compared to all possible factors that affect a traveler's behavior. These factors can be one factor that changes over time or many factors that the agent cannot fully observe.

The Azure personalizer was selected as it met all requirements and provided the ability to address the multi-step problem at scale using a **contextual bandits** reinforcement learning approach to determine the best action for a given context across all users to find the maximum average reward (Microsoft Learn 2022c). The contextual bandit learning algorithm is an extension of the multi-armed bandit approach that can test out different actions and automatically learn which action is most rewarding for a given context.

The personalizer needs the following information to determine the maximum average reward (Microsoft Learn 2022e):

- **Context:** Information that describes the current state of the application, e.g., the location of a job, the job type, and vice-versa the desired state to work in of traveler, and their certifications.

- **Actions:** a discrete set of items along with their attributes, e.g., a set of traveler recommendations for each job or a set of job recommendations for each traveler. It takes a limited set of actions, with approximately fifty actions in each rank API call. Since the is the possibility of getting more than fifty actions at a time, the smart match scoring system is used as a filtering mechanism.

- **The Reward:** a score between 0 and 1, where 0 is a bad choice, and 1 is a good choice.

Then this information was used to take advantage of the reinforcement learning system by using two APIs: **rank API** and **reward API** (Microsoft Learn 2022e). To train the personalizer models, HTTP-triggered and timer-triggered serverless functions pass a JSON file containing a set of actions, features to describe each action, and features to describe the current context.

Each call to the rank API is an event, with a unique identifier sent along. The personalizer then sends the best action with the event id and a probability that is the total average reward determined by the personalizer. The user can see the probability that the model has determined and can change it. The user's probability score change request calls the HTTP-triggered serverless function Reward endpoint, passing in the user's determined score along with the event. The user's feedback can help the personalizer learn if the new probability value is sent back. This HTTP-triggered serverless function trains the personalizer in an online mode as it learns from the feedback probability value provided by the user.

The timer-triggered serverless function runs once a week and acts as a batch learning system to train the bi-directional personalizers. Instead of a user returning a probability reward score, the batch system would use the smart match scoring plus some earlier user feedback to train the system. The batch learning method is also used to train the models in an initial ***Apprentice mode***, which helps it avoid a cold start problem with a new untrained model (Microsoft Learn 2022e). The personalizer uses collected information across all users to learn the best overall action based on current contexts (Microsoft Learn 2022e).

Figure 21 shows the Rank API call for traveler recommendations for a job. The job context feature contains job related features, the action features contains a list of traveler features, and the unique event id. The feature in job context, and traveler actions marked in red are excluded properties from the personalizer training, since they are not relevant to the model and are just there as a way to identify the probability scores returned by the personalizer response.

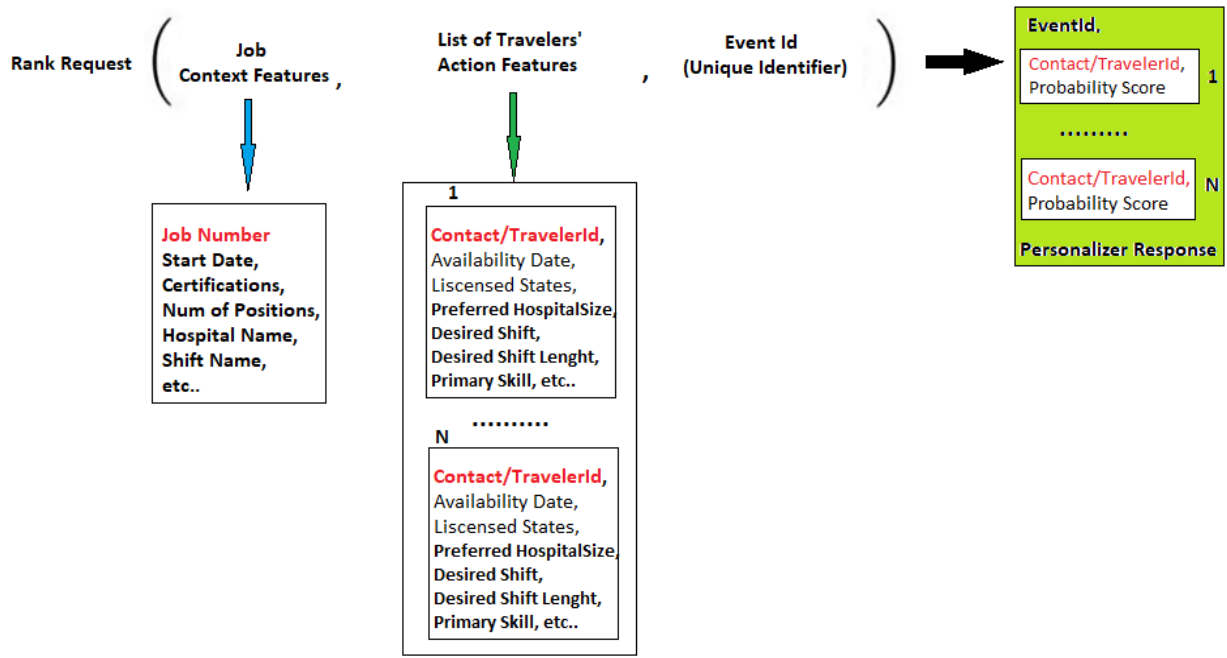

**Figure 21 –** Rank Request call to personalizer for traveler recommendations

Figure 22 show the inverse of figure 22 and calls the job recommendation for travelers. The context features has data for a traveler, a list of smart matched scored jobs for that traveler, and a unique number for the event id. The personalizer response returns the event id, and the list of jobs containing job number, and the probability score.

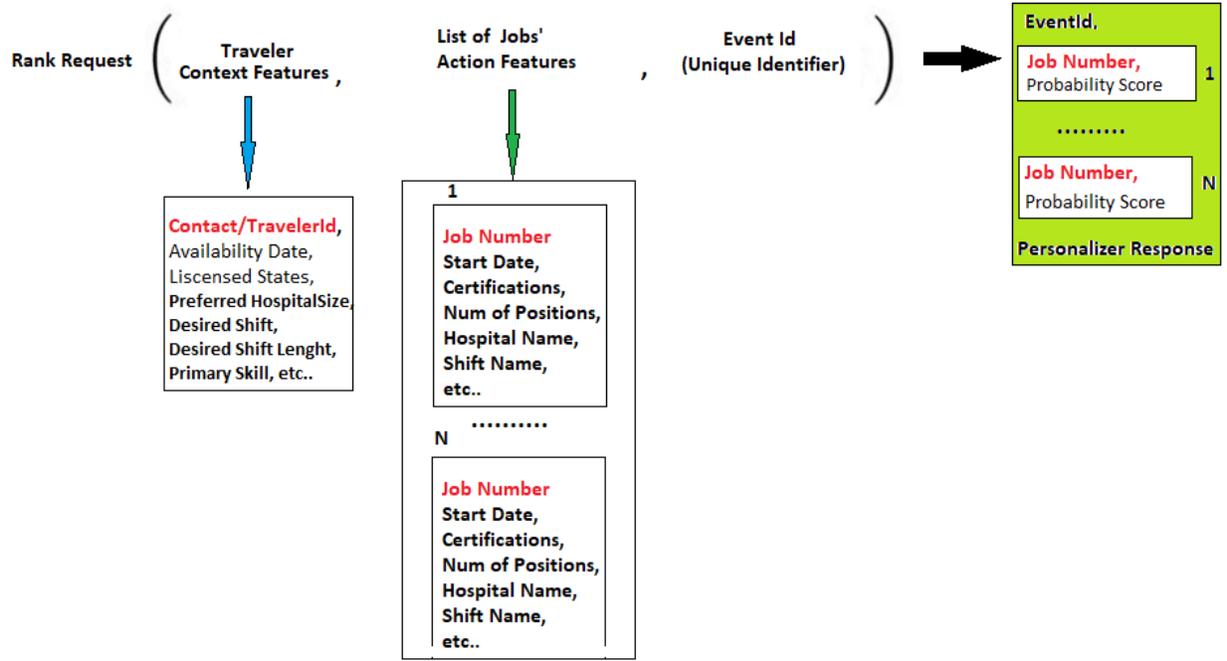

**Figure 22 –** Rank Request call to personalizer for job recommendations

Figure 23 shows the reward API call with the event id being the first argument. The value can depend on whether the input is from a batch learning or user-initiated interaction. If the user initiates the interaction, their returned probability score is used. The total smart match score is used as feedback if the system is in batch learning mode. This methodology is used for both the traveler recommendation for jobs and job recommendation for travelers' personalization models.

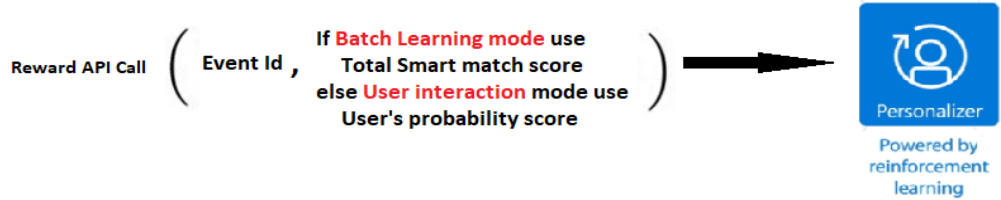

**Figure 23 –** Reward API call for the two personalization models

### 4.4.1 Active learning

Figure 24 shows the architecture of the active learning methodology initiated by user interactions. For example, the user can be a recruiter searching personalized active travelers for jobs or active jobs for travelers. Both models use the same architecture but the difference in the information context features and the actions as shown in figures 21 and 22. Since users are actively calling their appropriate personalization models, the top twenty actions with the highest total smart match scores are sent to the personalizer, and the scores from the personalizer are returned for the user to provide feedback. If there are more than twenty actions, and the user does not find what they are looking for, the next twenty actions are sent to the personalizer and then to the user for active feedback. The user returns their score for each action with the help of the reward API call as demonstrated by figure 23.

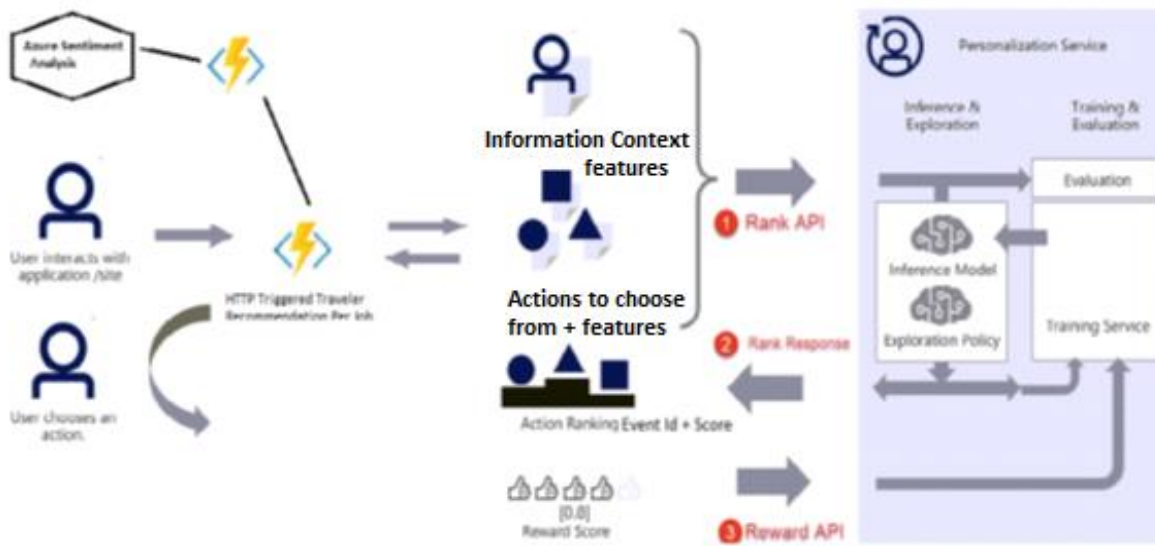

**Figure 24 –** User initiated active learning architecture for the two personalizers

Figures 25 and 26 show the traveler recommendation for each job and the job recommendations for each traveler's endpoints, along with their respective reward API calls.

```csharp
[FunctionName("TravelerRecommendation")]
2 references
public async Task<IList<Recommendation>> TravelerRecommendation(
    [HttpTrigger(AuthorizationLevel.Anonymous, "post", Route = null)] HttpRequest req,
    ILogger log)
{
    try
    {
        string requestBody = await new StreamReader(req.Body).ReadToEndAsync();
        var jobContext = JsonConvert.DeserializeObject<JobContextQuery>(requestBody);

        var jobRecommendations = await _service.GetRecommendations(jobContext);
        return jobRecommendations;
    }
    catch (Exception ex)
    {

    }
}

[FunctionName("Reward")]
0 references
public async Task Reward(
    [HttpTrigger(AuthorizationLevel.Anonymous, "post", Route = null)] HttpRequest req,
    ILogger log)
{
    try
    {
        string requestBody = await new StreamReader(req.Body).ReadToEndAsync();
        var reward = JsonConvert.DeserializeObject<Reward>(requestBody);

        await _service.Reward(reward);
    }
    catch (Exception ex)
    {

    }
}
```

**Figure 25** – Traveler recommendation per job endpoint

```
[FunctionName("JobRecommendation")]
2 references
public async Task<IList<Recommendation>> JobRecommendation(
    [HttpTrigger(AuthorizationLevel.Anonymous, "post", Route = null)] HttpRequest req,
    ILogger log)
{
    try
    {
        string requestBody = await new StreamReader(req.Body).ReadToEndAsync();
        var query = JsonConvert.DeserializeObject<TravelerContextQuery>(requestBody);

        var recommendations = await _service.GetJobRecommendations(query);
        return recommendations;
    }
    catch (Exception ex)
    {

    }
}

[FunctionName("TravelerReward")]
0 references
public async Task TravelerReward(
    [HttpTrigger(AuthorizationLevel.Anonymous, "post", Route = null)] HttpRequest req,
    ILogger log)
{
    try
    {
        string requestBody = await new StreamReader(req.Body).ReadToEndAsync();
        var reward = JsonConvert.DeserializeObject<Reward>(requestBody);

        await _service.TravelerReward(reward);
    }
    catch (Exception ex)
```

**Figure 26** – Job recommendation per traveler endpoint

## 4.4.2 Batch learning

Figure 27 shows the architecture of the batch learning methodology initiated by a weekly timer. The feedback reward score is provided by the smart match scoring for each model using their respective endpoints. The difference from the active learning endpoints is that context and actions are chosen randomly, with action lists being sent fifty at a time since this is the advice given by the documentation (Microsoft Learn 2022f). Figures 28 and 29 show the endpoints for the traveler and job recommendations, and they are both triggered to run once every Saturday.

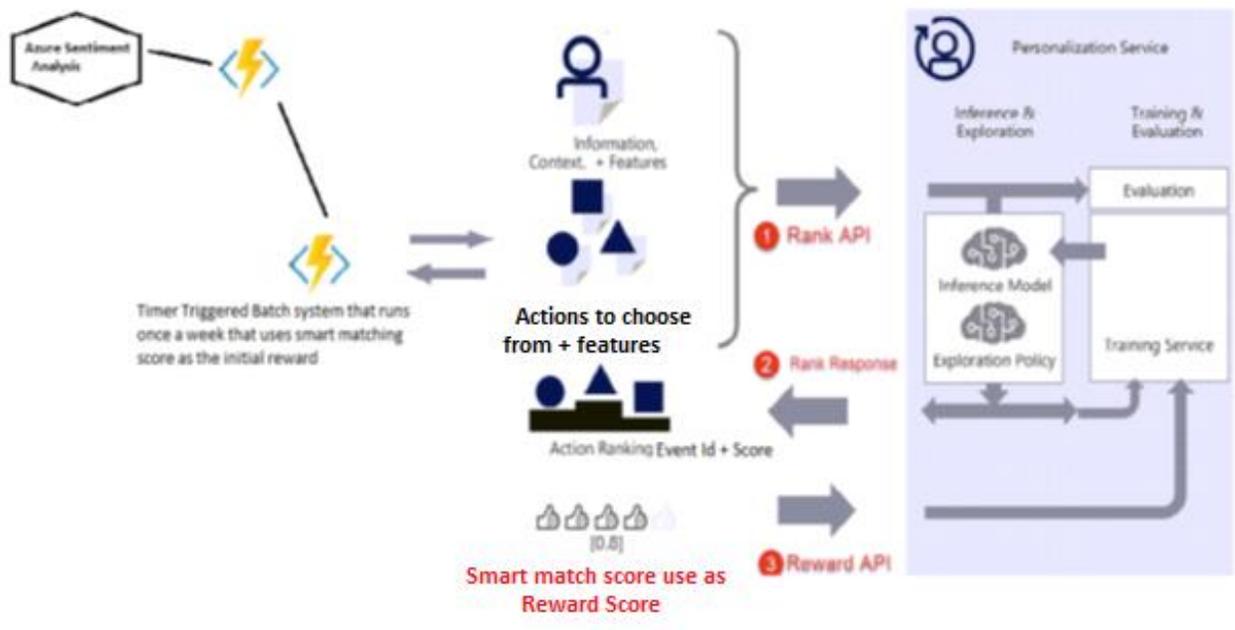

**Figure 27** – Batch learning architecture initiated with a weekly timer

```csharp
[FunctionName("BatchTrainTravelerRecommendations")]
1 reference | 0 changes | 0 authors, 0 changes
public async Task Run([TimerTrigger("0 0 0 * * SAT"
    )] TimerInfo myTimer, ILogger log)
{
    try
    {
        //Run once every saturday 0 0 0 * * SAT
        log.LogInformation($"{FunctionPrefix}: Random Batch Traveler Recommendations Training started.");

        await _service.BatchTrainTravelerRecommendationsAsync();

        log.LogInformation($"{FunctionPrefix}: Random Batch Traveler Recommendations Training completed.");
    }
    catch (Exception ex)
    {
        log.LogError(ex, $"{FunctionPrefix}: The function encountered an error.");
        throw;
    }
}
```

**Figure 28** – Traveler recommendation batch learning timer triggered API

```csharp
[FunctionName("BatchTrainJobRecommendations")]
1 reference | 0 changes | 0 authors, 0 changes
public async Task Run([TimerTrigger("0 0 0 * * SAT")] TimerInfo myTimer, ILogger log)
{
    try
    {
        //Run once every saturday 0 0 0 * * SAT
        log.LogInformation($"{FunctionPrefix}: Random Batch Job Recommendations Training started.");

        await _service.BatchTrainJobRecommendationsAsync();

        log.LogInformation($"{FunctionPrefix}: Random Batch Job Recommendations Training completed.");
    }
    catch (Exception ex)
    {
        log.LogError(ex, $"{FunctionPrefix}: The function encountered an error.");
        throw;
    }
}
```

**Figure 29** – Job recommendation batch learning timer triggered API

### Section 4.4.3 Feature engineering

A lot of time was spent trying to see what action features to add or remove, and when building the list, making sure that those features that contained null values or were not available were omitted from the request. For example, adding the *NonSerialized()* attribute to those properties ignores job numbers or traveler ids. Features were added and removed as training, and feature evaluation reports were run on each model, with examples shown in figures 30 and 31. These reports helped identify what features contributed to positive rewards and those less likely to contribute. Results were improved by seeing what features helped and the effects of adding or removing a feature.

**Feature effectiveness** ...

| 1 | j | availabilityDate | 92 | 510239 |
|---|---|---|---|---|
| 2 | j | PrimarySkillName | 84 | 630097 |
| 3 | j | AvailabilityDate | 78 | 510239 |
| 4 | j | primarySkillName | 70 | 630097 |
| 5 | j | LicensedStates | 69 | 7778 |
| 6 | j | careSetting | 62 | 599832 |
| 7 | JobProfiles | unitName | 61 | 49986 |
| 8 | j | shiftName | 60 | 599970 |

**Figure 30 –** Feature effectiveness report sample for traveler recommendations

**Feature effectiveness** ...

| Feature # | Namespace | Feature name | Score | Feature occurrences |
|---|---|---|---|---|
| 1 | j | BillRate | 100 | 20199 |
| 2 | j | NumberOfBeds | 74 | 20220 |
| 3 | j | ShiftName | 73 | 20220 |
| 4 | j | ParentAccountIdName | 72 | 16912 |
| 5 | j | HospitalTakeHomePay | 70 | 20199 |
| 6 | j | Certifications | 60 | 20204 |
| 7 | j | ClientLevel | 49 | 20220 |
| 8 | ContextFeatures | ShiftName | 47 | 565 |
| 9 | j | StateName | 47 | 20220 |

**Figure 31 –** Feature effectiveness report sample for job recommendations

Another model training setup was working through the rewards configurations, working with the exploration versus using learned models as shown in figure 32, where the default ten minutes sufficed to wait for reward API response time. The default reward was initially set to 0.5. Still, the model did not work as well since the reward API considered the event id and the reward value for the reward id ( the top performing id, which could be the traveler id for the traveler recommendation or the job id for the job recommendation system). The reward aggregation performed better, with reward aggregation being set to the earliest reward instead of the sum value, which was an aggregation of the reward values received.

**Figure 32 –** Reward wait time and aggregation setup

Additionally, the model update frequency settings took a lot of manipulation to get the best outcome for the model. The batch setup went from updating the model every ten minutes to every ten seconds for the initial training, as shown in figure 33. Once the model's cold start was overcome, the model update frequency value was set to ten minutes. Another lever used to manipulate the models was the data retention setup, where the two hundred days setting helped get enough data for the offline evaluations. It was used to measure the models' performances and help optimize their learning policies. The data retention setting is shown in figure 34, with the offline evaluations and model optimizations detailed in the next section.

**Figure 33 –** Model update frequency settings for the models

**Figure 34 –** Data retention setting for the models

## Section 4.4.4 Offline Evaluation and Model Optimization

An offline evaluation is available to evaluate, test and access the personalizers' effectiveness without downtime or changing code. The data management settings from figure 36 and a certain amount of rank and reward calls to the service help provide the following answers:

1. How effective are the personalizer ranks.

2. Which features of the context and actions are most effective or ineffective? Offline evaluations are in addition to the feature effectiveness reports described and shown in figures 31 and 32.

The Offline evaluation is used to find more optimized learning policies that the personalizers use to improve future results. To give successful recommendations, the models must have at least 50,000 events, and the data contain periodic and representative user behaviors and traffic. In addition, it is affected by the data retention setting shown in the previous section. In figure 35, you can see the various learning policies on the chart, their estimated average reward, confidence intervals, and options to download or apply a specific policy for the traveler recommendation model.

- "*Online*" – Personalizer's current policy (Microsoft Learn 2022a).

- "*Baseline1*" – Your application's baseline policy (Microsoft Learn 2022a).

- "*BaselineRand*" – A policy of taking actions at random (Microsoft Learn 2022a).

- "*Inter-len#*" or "*Hyper#*" – Policies created by Optimization discovery (Microsoft Learn 2022a).

You can then select Apply to select the policy that improves the model best for the available data.

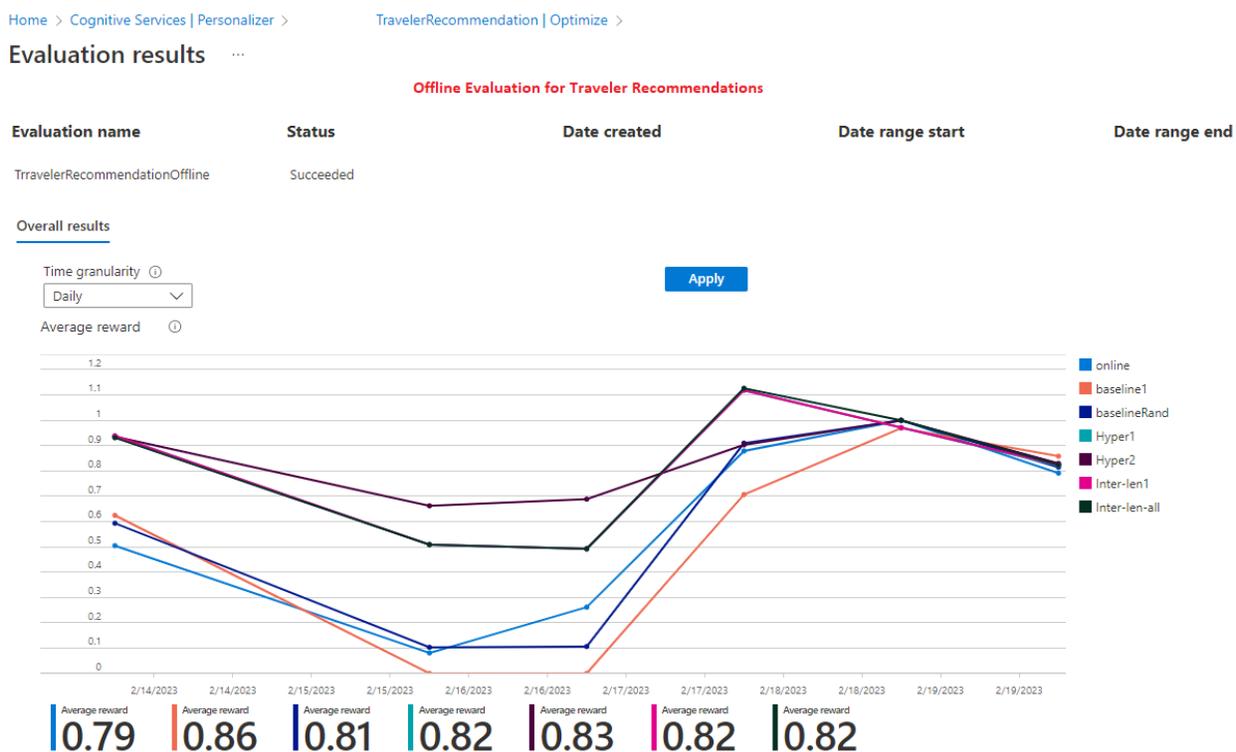

**Figure 35** – Offline evaluation report for traveler recommendations

Offline evaluations on both personalizers are being used to automatically find more optimal learning policies because it allows you to compare the current online policies to new suggested policies. Then, the new policies can be applied to the corresponding personalizers. Offline evaluations use a method called **Counterfactual Evaluation** only using observed user behavior.

Counterfactual Evaluations are grounded on the assumption that the rewards based on users' behaviors cannot predict what would have happened. The personalizers have no way of knowing what would have happened if the user had been presented with different data than what they have seen and, therefore, can only learn from rewards that have been given ( Microsoft Learn 2022a).

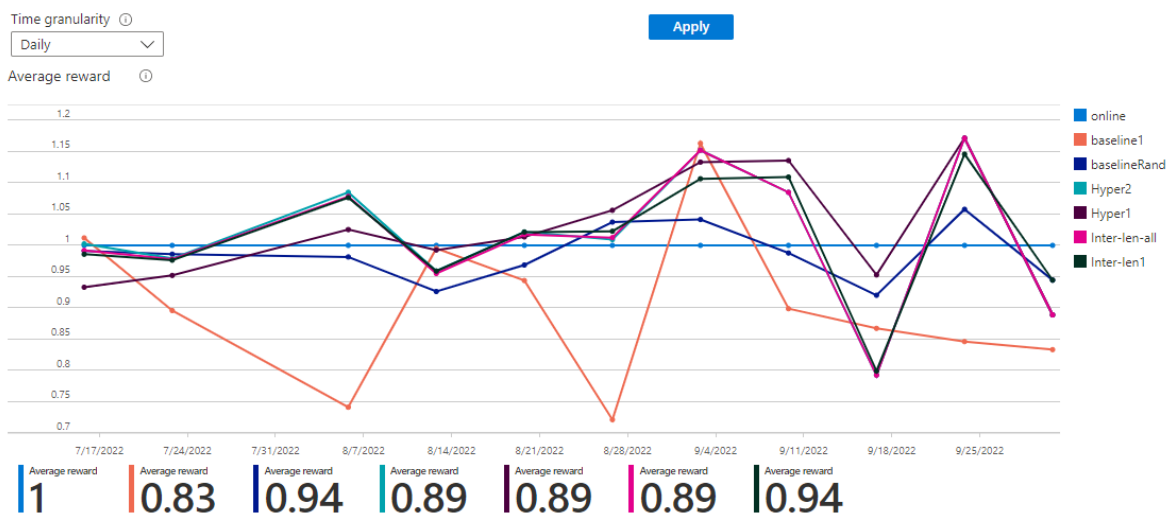

**Figure 36** – Auto-Optimize Offline 50valuation

Both personalizers have the auto-optimize setting turned on by default shown by figure 36. The auto-optimize setting saves a lot of manual effort on the personalizer loop at its best performance by automatically applying searched learning settings.

### Section 4.5 Test and model validation framework

The design of a machine learning system requires a way to set up a testing and validation framework to ensure that it achieves planned results. A testing and validation framework that ensures the model is tested robustly addresses challenges while considering testing and validation data. Continuous assessment of the system's performance demonstrates to stakeholders that it performs according to their expectations. Testing infrastructure for unit and integration tests that are compared and validated against copies of real production data helped conduct the assessments. During these steps, the NER mentioned earlier, personalization models are validated, data and model drifts are monitored, and fairness concerns are addressed.

### Section 4.6 Fairness

Fighting fairness (also referred to as bias) has been a topic at the forefront of machine learning, where fairness is seen as the absence of bias and vice versa (Chen et al. 2021). In the ML community, a common definition of fairness does not exist, where some have some definition of fairness with regards to the individual, and others to the group (Chen et al. 2021). Commonly mentioned sources of bias as listed below have been addressed in developing this system::

1. **Sampling bias:** this has to do with the process of collecting data, and issues of racism can manifest themselves at either the level of the individual or at a systemic level (Chen et al. 2021). The smart match and data collection process for our real-time Job Feed Processing system and Personalization do not take race or gender into account.

2. **Disparate treatment:** Bias can result in certain individuals receiving preferential treatment. For example, different genders are being treated differently (Chen et al. 2021). As mentioned before, the smart match system does not take gender or race into account. Additionally, the system does not consider what school they attended and only takes into account education level, certifications, licenses, etc.

3. **Tyranny of the majority:** The numerical majority affects the training loss the most through the sheer total numbers. In the personalization system, a dedicated recruiter could spend time unconsciously reevaluating the reward value of the reinforcement learning system, which has active learning, thereby ensuring that their travelers get most of the job recommendations. To consider this factor, a weekly batch training system acts as a buttress. As mentioned earlier, the data in a production (including actual matches, actual corrections, etc.) are copied every 15 minutes down to the lower environments that are used to validate the matches.

Additionally, reports can be generated, and questions can be asked for issues such as fairness and addressed as soon as possible.

# Chapter 5. Results

The results shown in the figures below are examples of the outputs at various stages of the personalization system. The first two parts consist of the SQL data that will be fed into the reinforcement learning personalization system. Figure 37 shows the SQL of the count of travelers that have matched some jobs using the smart match system. Figure 38 shows the number of jobs that were smart-matched individual travelers. This number of choices in figure 40 shows the need for personalizers since the smart match scoring filter shows that travelers have around five hundred seventy-three choices which would cause "decision paralysis".

| | JobId | TravelerMatchRecommendations |
|---|---|---|
| 1 | C70351A4- | 23 |
| 2 | B01B26DA- | 20 |
| 3 | F527314A- | 20 |
| 4 | 37665630- | 20 |
| 5 | 9C343C92- | 19 |
| 6 | 29FD6F89 | 19 |
| 7 | AE6C3A10- | 19 |
| 8 | 4F5968C7 | 19 |
| 9 | 00722680- | 19 |
| 10 | 08253679 | 19 |
| 11 | 30273462 | 19 |
| 12 | 4AE1DC95- | 19 |
| 13 | 3E629451 | 18 |

**Figure 37 –** Sample Count of Traveler Recommendations per Job Id

| | ContactID | JobRecommendationPerTraveler |
|---|---|---|
| 1 | 41EE9CC0 | 573 |
| 2 | DA6676B3- | 573 |
| 3 | 78347760- | 437 |
| 4 | 4F025FFE | 413 |
| 5 | 5AE9D601 | 309 |
| 6 | 3CB0E57 | 299 |
| 7 | 11655DC | 284 |
| 8 | 3C086BD | 265 |
| 9 | A8A83EC | 255 |
| 10 | 9FBC93E | 229 |

**Figure 38 –** Sample Count of job recommendations per traveler contact Id

Figure 39 is the log table that contains the job number, traveler number, event id, and probability returned for that row by the traveler recommendation personalization engine, the user-entered probability during the active learning phase, and the total smart match score. The observation for the traveler recommendations per job is shown below. The users returned a score closer to the smart match scoring, while the model was not initially performing as desired, as shown by the values in the probability column. The recommendations that have the user entered probability column with null values are those the user has not given feedback. Instead, it relies on the personalizer models' probability scores during batch learning.

| JobId | ContactId | EventId | Probability | UserEnteredProbability | TotalSmartMatchScore | PositionOpen |
|-------|-----------|---------|-------------|------------------------|----------------------|--------------|
| EA18A5F9 | 156C0496 | 6c1935df-b973-4830-9397-cc88ca0eee9f | 0.02 | 1.00 | 0.900000 | 2 |
| EA18A5F9 | 4DF64E00 | 5ab48320-1034-4595-9eee-c0f8b110494c | 0.02 | 1.00 | 0.900000 | 2 |
| EA18A5F9 | 6A576B6F | ff16e260-6606-4a34-b2b0f26ea428acda | 0.52 | 0.80 | 0.930000 | 2 |
| EA18A5F9 | 838EF867 | 2b7cc21c-40fd-4ab5-bb95-89156ffaab11 | 0.12 | 0.90 | 0.880000 | 2 |
| EA18A5F9 | ACB0A44E | 418f43c4-5bdc-4c03-9e5a-f770c8db32c0 | 0.02 | 0.80 | 0.880000 | 2 |
| EA18A5F9 | C73778E9 | ebc6a5a9-e3e2-4390-a745-aa77e1bff94a | 0.75 | 1.00 | 0.930000 | 2 |
| EA18A5F9 | 3CB0E578 | bde86897-754b-40bb-b02e-539ad9b7f528 | 1.00 | 1.00 | 0.930000 | 2 |
| EA18A5F9 | 78347760 | a2acb79f-c178-4aa7-a493-a5ba9373577a | 0.67 | 1.00 | 0.900000 | 2 |
| 4C43FFDA | 54C43192 | ec60cc904955-44c8-9f3b-ee0d1200c766 | 0.12 | 0.80 | 0.900000 | 1 |
| 4C43FFDA | 7D17355A | bfd9a850-3daf-4a36-89ec-eb7beb0d17be | 0.06 | 1.00 | 0.870000 | 1 |
| 4C43FFDA | E7362125 | 60a9f85e-8fc2-4182-ae2c-ed0ea5044ff3 | 0.67 | NULL | 0.900000 | 1 |
| 4C43FFDA | 4ED29AFE | f541bdc3-4061-40c2-a294-4b61be0dbe80 | 1.00 | NULL | 0.920000 | 1 |
| 4C43FFDA | DC04D5D1 | 405463a1-ee63-4877-9f64-e86d4bcae773 | 0.10 | NULL | 0.900000 | 1 |
| 4C43FFDA | 7D6081D7 | bb3eecbb-8a20-4b9e-aad4-4952bb5cbd25 | 0.08 | NULL | 0.900000 | 1 |
| 4C43FFDA | 208B52C3 | 11c2f916-b96d-4b96-a86b-9b621eb3950f | 0.75 | NULL | 0.900000 | 1 |
| 4C43FFDA | A28D834B | 09814651-9e17-40d8-a202-15889478b1ce | 0.57 | NULL | 0.870000 | 1 |

**Figure 39** – Traveler personalizer recommendation models initial run sample

| JobId | ContactId | EventId | Probability | UserEnteredProbability | TotalSmartMatchScore | PositionOpen |
|-------|-----------|---------|-------------|------------------------|----------------------|--------------|
| EA18A5F9 | 156C0496 | 0E93BA44-12BF-4DFA-8023-D299A3038D0A | 0.86 | 1.00 | 0.900000 | 2 |
| EA18A5F9 | 4DF64E00 | 1593B928-2227-432D-B528-63C3BF121ADE | 0.94 | 1.00 | 0.900000 | 2 |
| EA18A5F9 | 6A576B6F | A324C7E9-C438-4379-A116-4EE2B2879921 | 0.94 | 0.80 | 0.930000 | 2 |
| EA18A5F9 | 838EF867 | 021517B7-3C26-45E1-81AC-AAD9D55185B0 | 0.86 | 0.90 | 0.880000 | 2 |
| EA18A5F9 | ACB0A44E | 85774B2E-85F7-4306-9CC0-E14DFF998CE0 | 0.89 | 0.80 | 0.880000 | 2 |
| EA18A5F9 | C73778E9 | 8A405BED-0095-4CBE-B296-1A16E734E92A | 0.90 | 1.00 | 0.930000 | 2 |
| EA18A5F9 | 3CB0E578 | 66A56898-272D-4BDB-998A-C00205C22BCD | 0.90 | 1.00 | 0.930000 | 2 |
| EA18A5F9 | 78347760 | 42EF0401-5065-443F-AFED-13DC8ADBA025 | 0.97 | 1.00 | 0.900000 | 2 |
| 4C43FFDA | 54C43192 | 94AC77ED-1438-47EB-907A-78878B68258F | 0.94 | 0.80 | 0.900000 | 1 |
| 4C43FFDA | 7D17355A | 2A4435DA-8982-4353-94AF-0BA5DC5E390C | 0.96 | 1.00 | 0.870000 | 1 |
| 4C43FFDA | E7362125 | 078CC584-5601-4874-8703-D5AE2B322F16 | 1.00 | NULL | 0.900000 | 1 |
| 4C43FFDA | 4ED29AFE | 4F319A7D-0939-4C98-BFAF-319A2C686BA5 | 1.00 | NULL | 0.920000 | 1 |
| 4C43FFDA | DC04D5D1 | 427E19F9-003F-4F8D-97E0-522744F078DC | 1.00 | NULL | 0.900000 | 1 |
| 4C43FFDA | 7D6081D7 | 132E4D81-80A8-4902-9D69-82A0E21B85BD | 0.88 | NULL | 0.900000 | 1 |
| 4C43FFDA | 208B52C3 | F18ED568-8116-4901-8219-3BA9E2376D5D | 0.86 | NULL | 0.900000 | 1 |
| 4C43FFDA | A28D834B | 51791D5B-4B57-4DF9-9B39-884E8EA2F366 | 0.90 | NULL | 0.870000 | 1 |

**Figure 40** – Traveler personalizer recommendation models after twenty thousand runs sample

Figure 40 above shows the same results after twenty thousand random iterations with the traveler recommendation model's probability scores closely matching the user-entered or total smart match scores. Figure 41 shows a sample of the initial run of the job recommendation personalizer for travelers. A few model-generated probability scores show a close match to the total smart match score. Figure 41 shows after the twenty thousandth run, where we see some marked improvements. We can see that the highlighted row shows that when ordered by probability score, the score with an initial score of 0.75 now had a score of 0.89. The other scores closely matched the total smart match score.

| ContactId | JobId | EventId | Probability | UserEnteredProbability | TotalSmartMatchScore | PositionOpen |
|---|---|---|---|---|---|---|
| FFF3AB27 | 3B0F4027 | 58AC898C-D4E3-41AF-9B42-C79A5A0583D2 | 1.00 | NULL | 0.950000 | 1 |
| FFF3AB27 | 528FDAC1 | 9067F981-E79A-46A8-8375-385E4CA072F8 | 1.00 | NULL | 0.950000 | 3 |
| FFF3AB27 | 6902273F | 08C65705-9DCA-4E25-82B1-94E6BBD495EE | 0.75 | NULL | 0.920000 | 1 |
| FFF3AB27 | 1EB33C5B | A55BC686-8929-4E0F-B137-FFC76742D24E | 0.25 | NULL | 0.920000 | 1 |
| FFF3AB27 | 372A0C76 | A5761A45-027B-4879-9EAA-930DB76D9202 | 0.17 | NULL | 0.920000 | 1 |
| FFF3AB27 | CC7CD089 | 1531B9DF-FEB2-4521-BBB8-D7E01A31DE51 | 0.17 | NULL | 0.920000 | 1 |
| FFF3AB27 | 3F8B6F8A | 8375B779-8605-4E3E-A4A5-D574849AD0F6 | 0.07 | NULL | 0.900000 | 3 |
| FFF3AB27 | 64A9DE19 | 67932CA4-8774-4A3F-BBCD-2EFB226B66AF | 0.05 | NULL | 0.910000 | 1 |
| FFF3AB27 | F4B0EA94 | 7365D99B-45C1-4346-8B48-D6EF875B9A0C | 0.05 | NULL | 0.910000 | 2 |
| FFF37506 | 989F6092 | 424406C8-2961-43ED-8025-40CBC86C51BF | 1.00 | NULL | 0.900000 | 2 |
| FFF37506 | 6EDBE722 | AC680D42-9BAE-489A-8DEF-DE1680C599DC | 0.25 | NULL | 0.900000 | 8 |

**Figure 41** – Job personalizer recommendation models initial run sample

| ContactId | JobId | EventId | Probability | UserEnteredProbability | TotalSmartMatchScore | PositionOpen |
|---|---|---|---|---|---|---|
| FFF3AB27 | 3B0F4027 | C314464C-BA23-41EE-8FE7-1B9ABE477BCA | 1.00 | NULL | 0.950000 | 1 |
| FFF3AB27 | 528FDAC1 | A0D67E73-BA3B-4756-AB49-92F30F043497 | 0.97 | NULL | 0.950000 | 3 |
| FFF3AB27 | 372A0C76 | B7F868B8-A2D6-4451-A18F-C7600D9BEEFB | 0.94 | NULL | 0.920000 | 1 |
| FFF3AB27 | CC7CD089 | E8FDA8E2-6853-42AC-AEAB-F43BC8EF8FFB | 0.94 | NULL | 0.920000 | 1 |
| FFF3AB27 | 1EB33C5B | FC96FDEA-BA88-4942-9ED1-5E640885F0AA | 0.90 | NULL | 0.920000 | 1 |
| FFF3AB27 | 3F8B6F8A | 6AFC25DE-D16C-4D9A-B517-7594C8EB8517 | 0.90 | NULL | 0.900000 | 3 |
| FFF3AB27 | 64A9DE19 | A5880785-109B-4F77-9EC0-10DE05956C46 | 0.90 | NULL | 0.910000 | 1 |
| FFF3AB27 | F4B0EA94 | 3CF3CFD4-129D-48BE-B736-805BB3A44C9C | 0.90 | NULL | 0.910000 | 2 |
| FFF3AB27 | 6902273F | 921408BF-02B7-4B8B-B81C-3D05832FF4BF | 0.89 | NULL | 0.920000 | 1 |

**Figure 42** – Job personalizer recommendation models after twenty thousand runs sample

Suppose we wanted to see the sum of successful calls to the traveler or job recommendation models within a specific timeline. In that case, we could generate the reports in figures 43 and 44.

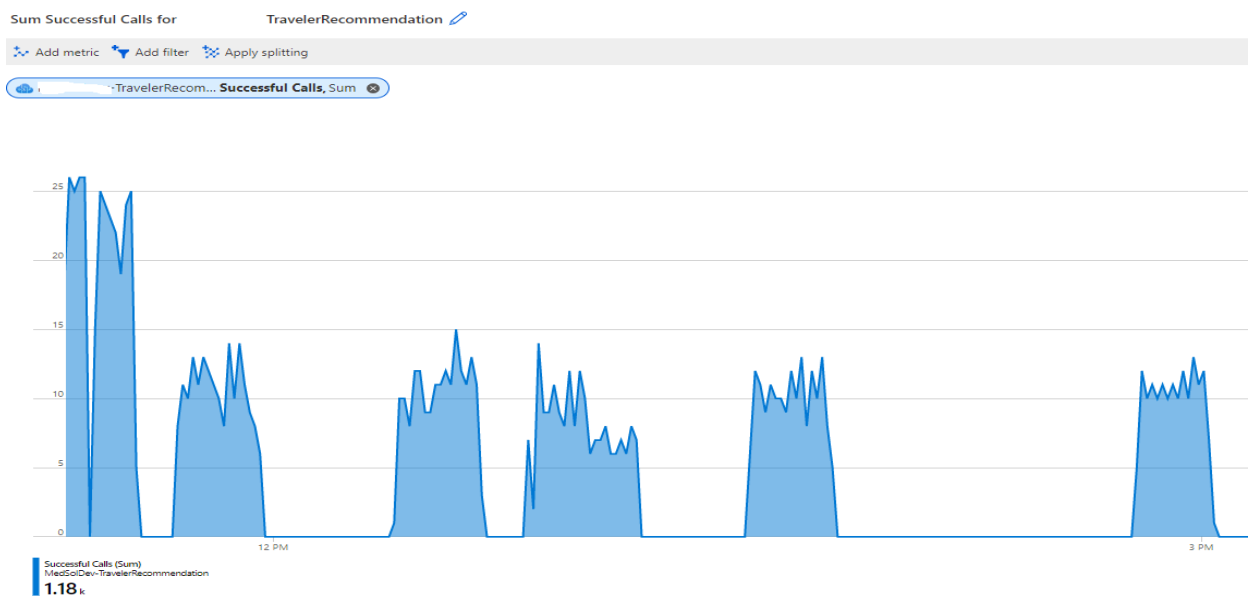

**Figure 43 –** Sum of successful calls to the traveler recommendation model with a timeline

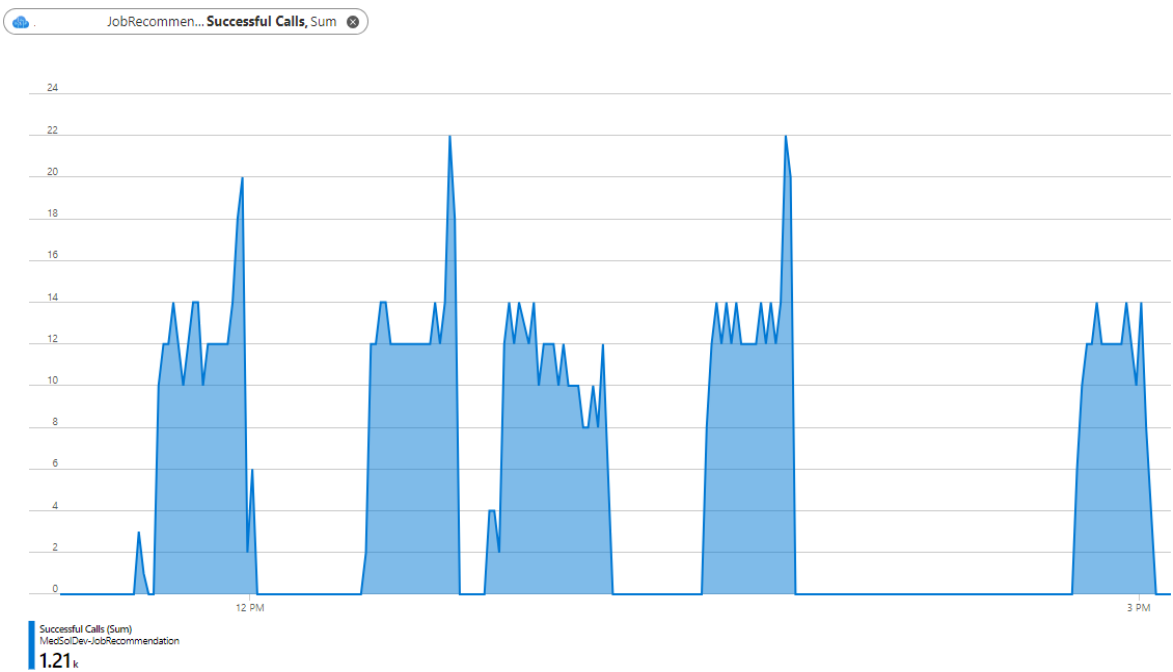

**Figure 44 –** Sum of successful calls to the job recommendation model with a timeline

The availability of feature evaluation reports, offline and auto-optimize models that can be applied to the models and their reports provide many opportunities to be solid personalizers that alleviate "decision paralysis" and give better-personalized recommendations tailored to users when built on top of the smart match scoring filtration. Additionally, more context and action features can be considered through offline evaluations and feature reports to help the model better understand its environment.

## Chapter 6. Conclusions

Although this system is a good starting point, the ever-evolving nature of the business requires the system to continue to build trust to increase user engagement. Having users trust the recommendations will eventually maximize positive behavior change and increase response time to staffing jobs and productivity. Additionally, machine learning-specific failures like production data differing from training data have been addressed in this design cycle since data is copied from production databases to the other database's environments (Huyen 2022, 361). Eventually, the small number of edge cases should be found and addressed, as they could cause the model to make catastrophic mistakes (Huyen 2022, 365).

The data service uses an orchestration of natural language processing (NLP) models that efficiently synthesize job-related data into a database and accurately cut down new or updated job acquisitions from a few hours, a week, or worse, to a few minutes. The data service improvements would also improve the personalized recommendations of the bi-directional personalization systems. For example, offline evaluations, auto-optimization, and feature reports have all helped improve the performance of the personalizer systems and enabled them to give a maximum of twenty recommendations per job and its corresponding openings in a near real-time manner.

## Chapter 7. Future work

For the next phase of improving the personalized recommendations, the reward system should include hiring numbers scores instead of just submission numbers. Additionally, since submission numbers are currently the lead indicator for hiring numbers, the personalizers should improve to the point that they return the maximum of the top five recommendations per job or traveler instead of the top twenty. Then the results of the changes can be compared to see if they improved the hiring numbers.

The next work to improve the personalizers would be to toggle between the exploration versus the learned features settings to see what further improvements the models can make. For example, the exploration feature would choose a different action for a user instead of just the best action and provide a better-personalized recommendation by adapting to the ongoing user behavior through exploration ( Microsoft Learn 2022b*).*

The Bi-directional personalized recommendations with active learning are the kind of systems that could be ported to other industries whose industry requires inverse systems with active feedback, not just the medical staffing industry. An example is every recruiting service or company that needs to match jobs and potential hires. These bi-directional systems can use each company's job and applicant custom matching system with minor tweaks to context and action features. This would cut down on time to fill a company's open positions while ensuring applicants get as many top N recommendations as possible.

# Appendix A

## A content based traveler recommender based on TF-IDF and Cosine similarity

Appendix A shows an example of a content-based traveler recommendation based on traveler skills matching a job's skill needs. Recommendations can be based on content similarity and get started with two natural language processing (NLP) techniques: Term-Frequency Inverse Document Frequency (TF-IDF) and cosine similarity (Nixon 2022). A traveler recommender based on job skill categories first uses a TF-IDF to obtain a vector representation of the data. Figure A.1 shows the TF-IDF formula that will help capture the recommended skill for each traveler by giving a higher weight to the less frequent skills.

$$w_{i,j} = tf_{i,j} \cdot log\left(\frac{N}{df_i}\right)$$

$$tf_{i,j} = number\ of\ occurences\ of\ i\ in\ j$$

$$df_{i,j} = number\ of\ documents$$

$$N = total\ number\ of\ documents$$

**Figure A.1** – TF-IDF formula

Figure A.1 shows the product of the term frequency, i.e., the amount of times a given skill occurs in a skills of a traveler, times the right side term, which scales the term frequency depending on the number of times a given term appears in all travelers. The fewer travelers with the given skill ($df_i$), the larger the weight. The log is there to normalize the result of the division (Nixon 2022). Once the td-idf vectors are identified, cosine similarity, whose formula is shown in figure A.2, can define the closest matches. The lower the angle between the vectors, the higher the cosine value, yielding a higher similarity factor (Nixon 2022). The model takes the TF-IDF matrix and the cosine similarity scores for all the traveler skill documents and then generates content-based recommendations.

$$cosine\ similarity = \frac{r_i \cdot r_k}{|r_i||r_k|} = \frac{\sum_{j=1}^{m} r_{ij} r_{kj}}{\sqrt{\sum_{j=1}^{m} r_{ij}^2 \sum_{j=1}^{m} r_{kj}^2}}$$

**Figure A.2** – Cosine similarity formula